\documentclass[twocolumn,times]{aastex62}

\usepackage{color}
\usepackage{subfigure}
\usepackage{amsmath}
\usepackage{float}
\usepackage{lineno}
\usepackage{placeins}

\newcommand{\Deg}{^{\circ}} 
\submitjournal{\apj}
\shorttitle{Supergalactic Structure of Multiplets of UHECR}
\shortauthors{Abbasi et al.}

\begin{document}

\title{Evidence for a Supergalactic Structure of Magnetic Deflection Multiplets of Ultra-High Energy Cosmic Rays}

\correspondingauthor{J.P. Lundquist}
\email{jplundquist@gmail.com}

\author[0000-0001-6141-4205]{R.U. Abbasi}
\affiliation{Department of Physics, Loyola University Chicago, Chicago, Illinois, USA}

\author{M. Abe}
\affiliation{The Graduate School of Science and Engineering, Saitama University, Saitama, Saitama, Japan}

\author[0000-0001-5206-4223]{T. Abu-Zayyad}
\affiliation{High Energy Astrophysics Institute and Department of Physics and Astronomy, University of Utah, Salt Lake City, Utah, USA}

\author{M. Allen}
\affiliation{High Energy Astrophysics Institute and Department of Physics and Astronomy, University of Utah, Salt Lake City, Utah, USA}

\author{R. Azuma}
\affiliation{Graduate School of Science and Engineering, Tokyo Institute of Technology, Meguro, Tokyo, Japan}

\author{E. Barcikowski}
\affiliation{High Energy Astrophysics Institute and Department of Physics and Astronomy, University of Utah, Salt Lake City, Utah, USA}

\author{J.W. Belz}
\affiliation{High Energy Astrophysics Institute and Department of Physics and Astronomy, University of Utah, Salt Lake City, Utah, USA}

\author{D.R. Bergman}
\affiliation{High Energy Astrophysics Institute and Department of Physics and Astronomy, University of Utah, Salt Lake City, Utah, USA}

\author{S.A. Blake}
\affiliation{High Energy Astrophysics Institute and Department of Physics and Astronomy, University of Utah, Salt Lake City, Utah, USA}

\author{R. Cady}
\affiliation{High Energy Astrophysics Institute and Department of Physics and Astronomy, University of Utah, Salt Lake City, Utah, USA}

% ...

\begin{abstract}
Evidence for a large-scale supergalactic cosmic ray multiplet (arrival directions correlated with energy) structure is reported for ultra-high energy cosmic ray (UHECR) energies above 10$^{19}$~eV using seven years of data from the Telescope Array (TA) surface detector and updated to 10 years. Previous energy-position correlation studies have made assumptions regarding magnetic field shapes and strength, and UHECR composition. Here the assumption tested is that, since the supergalactic plane is a fit to the average matter density of the local Large Scale Structure (LSS), UHECR sources and intervening extragalactic magnetic fields are correlated with this plane. This supergalactic deflection hypothesis is tested by the entire field-of-view (FOV) behavior of the strength of intermediate-scale energy-angle correlations. These multiplets are measured in spherical cap section bins (wedges) of the FOV to account for coherent and random magnetic fields. The structure found is consistent with supergalactic deflection, the previously published energy spectrum anisotropy results of TA (the hotspot and coldspot), and toy-model simulations of a supergalactic magnetic sheet. The seven year data post-trial significance of this supergalactic structure of multiplets appearing by chance, on an isotropic sky, is found by Monte Carlo simulation to be 4.2$\sigma$. The ten years of data post-trial significance is 4.1$\sigma$. Furthermore, the starburst galaxy M82 is shown to be a possible source of the TA Hotspot, and an estimate of the supergalactic magnetic field using UHECR measurements is presented.
\end{abstract}

\keywords{astroparticle physics, cosmic rays, UHECR, supergalactic plane, multiplets, magnetic deflection, large-scale structure of universe}

\section{Introduction}\label{sec:intro}

The supergalactic plane (SGP) is the average matter distribution of the local universe up to a distance of $\sim$200~Mpc (a large percentage of its sources are within the GZK horizon of 100~Mpc) \cite{1975ApJ...202..610D}. Large scale magnetic fields have been measured between clusters of galaxies, which make up the supergalactic plane, including the Coma Cluster, and a $\sim$3~Mpc field between Abell 0399 and Abell 0401 (\cite{1975ApJ...202..610D}, \cite{2010A&A...513A..30B},  \cite{Govoni981}). 

It has also been shown that $\sim$90\% of the baryonic mass of the universe is contained between galaxies, of which $\sim$40\% is warm-hot protons outside gas clouds \cite{Nicastro2018}. This may support the formation of even larger intra-galactic scale magnetic fields (\cite{Biermann:1996xi}, \cite{Ryu:1998up}). The presence of large scale magnetic fields suggests that energy-dependent deflection of ultra-high energy cosmic rays (UHECR) may appear correlated with the SGP. 

Previous UHECR energy-position correlation (multiplet) searches for small scale galactic magnetic deflections have not had significant results (\cite{2012APh....35..354P}, \cite{2015EPJC...75..269A}, \cite{Bretz2011}, \cite{Wirtz:2019}). These multiplet searches used linear correlations of angular distance versus $1/energy$ and included scanned parameters chosen by assumed magnetic field models and compositions. The present analysis uses intermediate-scale energy-position correlations (multiplets) to look for significant large scale magnetic, and source, structure with minimal assumptions regarding particular magnetic field models or UHECR composition. 

In this paper the oversampled multiplets are found at grid points evenly covering the field-of-view sky (FOV), each having their own parameters of size, shape, pointing direction, and energy threshold. The structure of these multiplets is consistent with supergalactic deflection, the previously published energy spectrum anisotropy results of TA (the hotspot and coldspot) (\cite{Abbasi2018E}), and toy-model simulations of a supergalactic magnetic sheet \cite{Biermann:1996xi}. Here we report the significance using seven years of Telescope Array (TA) data (as in \cite{LundquistUHECR2019}) and update it to ten years of data.

\section{Energy-Angle Correlations}\label{sec:EDCorr}

\begin{figure}[htb]
\centering
\subfigure[]{%
   \includegraphics[width=.46\textwidth]{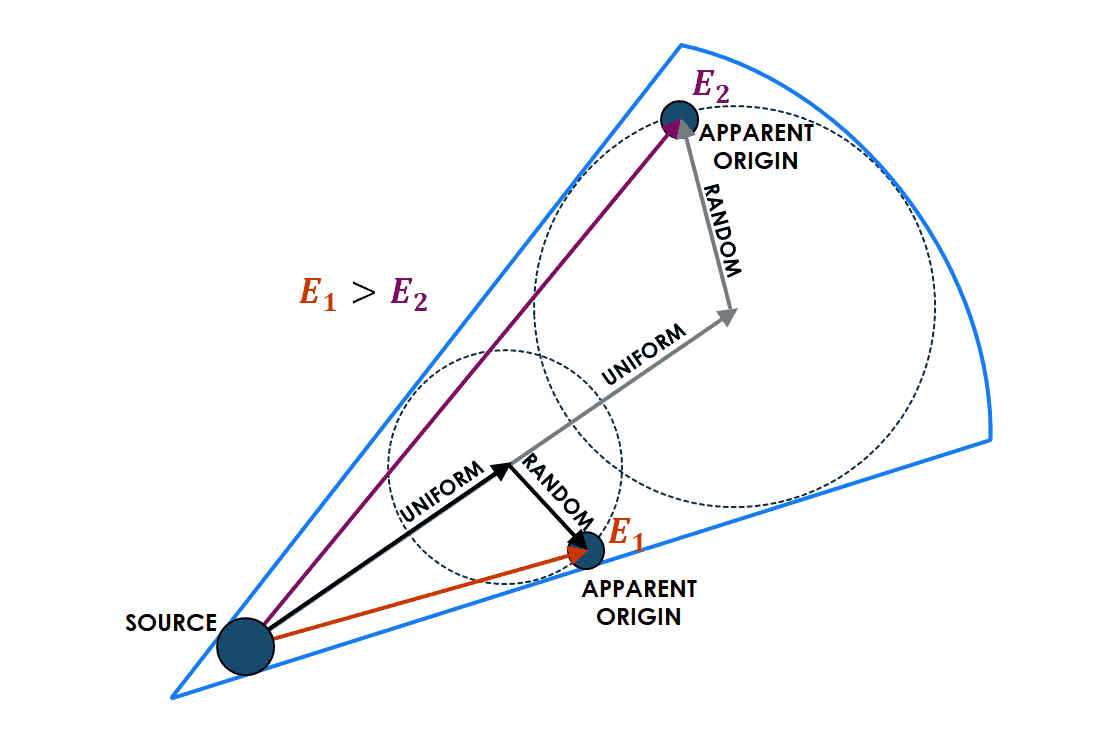}}
   \subfigure[]{%
   \includegraphics[width=.45\textwidth]{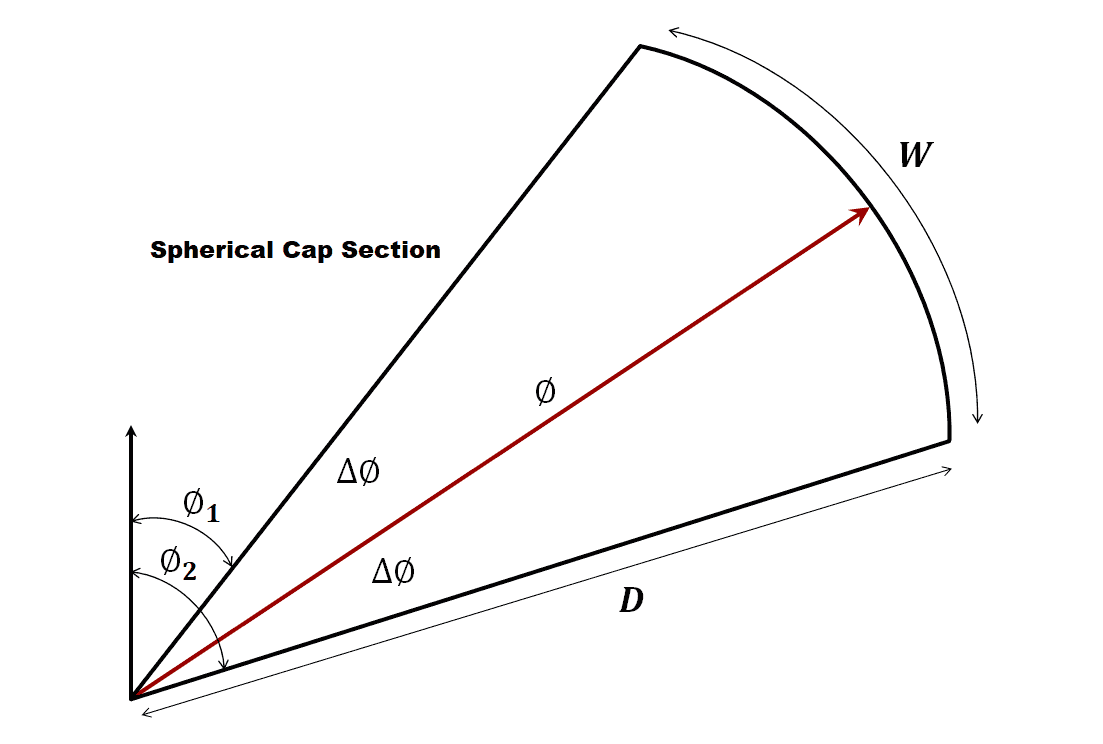}}
  \caption{Pictograph of UHECR Drift-diffusion deflection ``wedge'' bins (spherical cap sections) displayed on a flat space. (a) Two different energy events having traveled through coherent and random magnetic fields. The purple vector represents the low energy event spherical arc, and the red vector is a higher energy event. Coherent and random magnetic field components describe the average perpendicular to the field-of-view (FOV) sphere. Dashed circles represent possible random field RMS deflections. (b) A spherical cap section (wedge) is a simple shape that best encompasses the likeliest positions. Pointing direction is the spherical arc $\phi$, $\Delta \phi$ is the wedge width, and $D$ is the maximum angular distance (spherical cap radius).}\label{fig:deflection}
\end{figure}

It is assumed that UHECR are deflected as they travel through coherent magnetic fields according to Equation~\ref{eq:deflect}, with a deflection variance by random fields as approximated by Equation~\ref{eq:randdef} ($Z$ is mass number, $B$ is field strength, $S$ is distance traveled in the field, $E$ is particle energy, and $L_c$ is mean magnetic field coherence length). These deflection equations are from \cite{Roulet:2003rr} in units more relevant to the extragalactic case. The end effect of these fields is that lower energy cosmic ray events are deflected to larger angular distances from their source than higher energy events in both lateral and transverse directions \cite{Roulet:2003rr}. This drift-diffusion process is diagrammed in Figure~\ref{fig:deflection}.

\begin{subequations}
\begin{equation}
\delta \approx 0.5 \Deg Z\frac{B}{\mathrm{nG}}\frac{S}{\mathrm{Mpc}}\frac{10^{20} \mathrm{eV}}{E}
\label{eq:deflect}
\end{equation}
\begin{equation}
\delta_{rms} \approx 0.4\Deg Z \frac{B_{rms}}{\mathrm{nG}} \frac{10^{20} \mathrm{eV}}{E} \sqrt{\frac{S}{\mathrm{Mpc}}} \sqrt{\frac{L_c}{\mathrm{Mpc}}}
\label{eq:randdef}
\end{equation}\label{eq:bothdeflections}
\end{subequations}

\subsection{Correlation}
The distance between two points on the surface of a sphere, the great circle angular distance, is shown in Equation~\ref{eq:dist} in terms of vectors normal to the field-of-view. Correlations between event energy and angular distance from a grid point are found using a ranked correlation, Kendall's~$\tau$, that measures the strength of monotonic dependence \cite{KENDALL01111945}. 

\begin{equation}
\delta_{ij} = \arctan \frac{|\mathbf{\hat{n}_i} \times \mathbf{\hat{n}_j}|}{\mathbf{\hat{n}_i}\cdot \mathbf{\hat{n}_j}}
\label{eq:dist}
\end{equation}

The Kendall correlation is generally more robust against noise than the other common ranked correlation - Spearman's~$\rho$ \cite{Croux2010}. Ranked correlation minimizes the effects on correlation strength by magnetic model (such as higher-order terms of Equation~\ref{eq:deflect}), composition assumption, energy reconstruction systematics, and detector exposure variation.

Kendall's $\tau$ ranked correlation is the linear correlation between the separate ordering of the two variables of interest (variable $x$ sorted ranks: 1st, 2nd, 3rd, etc. versus variable $y$ ranks: 5th, 1st, 4th, etc.), with $n$ pairs of values, and is shown in Equation~\ref{eq:kendall}.

\begin{equation}
\tau = \frac{2}{n(n-1)}\sum_{j<k} sign \left[ \frac{x_{j}-x_{k}}{y_{j}-y_{k}} \right]
\label{eq:kendall}
\end{equation}

This correlation can be considered simply as the normalized sum of the sign of the slopes between all pairs of data points. A small correction is made for the rare occurrence of duplicate values and can be found in \cite{KENDALL01111945}.

The correlation coefficient $\tau$ has a range from $-1$ to $+1$, and a value of zero means that there is no measured relationship between the variables. For $+1$, an increase ($decrease$) of $x$ always follows an increase ($decrease$) of $y$. If $\tau = -1$ an increase ($decrease$) of $x$ always follows a decrease ($increase$) in $y$ (in this analysis x and y are energy and angular distance). A negative correlation is consistent with the expectation for magnetic field deflected events - as energy decreases, deflection increases, as can be seen from Equation~\ref{eq:bothdeflections}.

Any monotonic function ($x^{b}$, $log_{10}(x)$, $e^{x}$, etc.) of distance, energy, or both will always return a $\tau$ coefficient with the same magnitude but not necessarily the same sign ($\pm$). The sign of the resulting $\tau$ would be the original $\tau$ multiplied by the signs of the first derivatives of the applied functions.

The pre-trial two-sided significance of a correlation, $z$, (probability of $\tau$$=$$0$) is a function of correlation strength and sample size $n$. This significance is found by counting permutations of the sample ranks with greater $\tau$, or in the large $n$ sample limit, Equation~\ref{eq:corrsig} (for $n$$\geq$50), which follows the standard normal distribution. Further details can be found in \cite{KENDALL01111945}.

\begin{equation}
z = \frac{\tau 3 n(n-1)/2}{\sqrt{n(n-1)(2n+5)/2}}
\label{eq:corrsig}
\end{equation}

\subsection{Correlation Binning}\label{ssec:wedge}

\begin{figure}[tb]
    \includegraphics[width=1\linewidth]{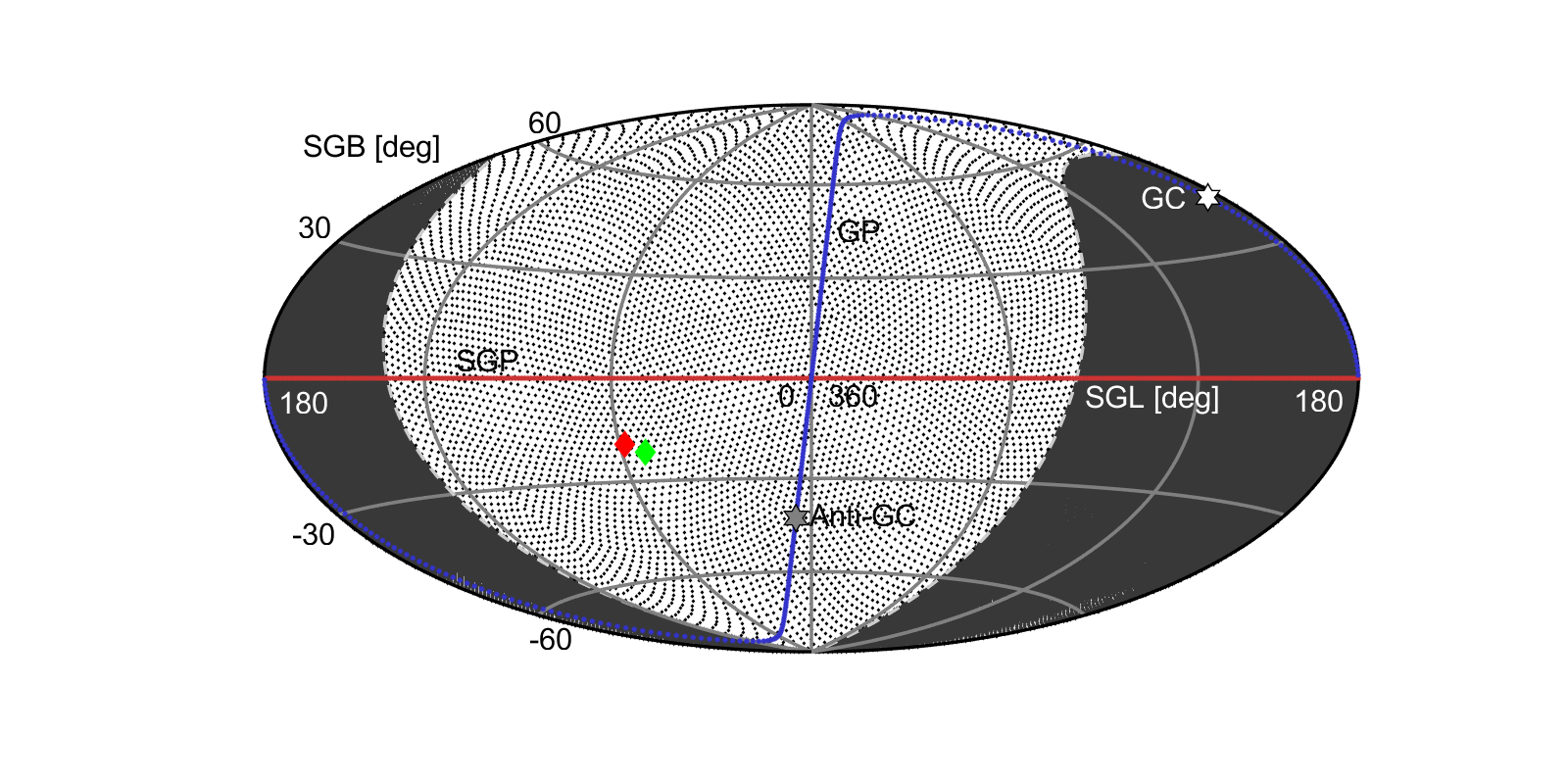}
  \caption{A supergalactic Hammer-Aitoff projection of the equal distance oversampling grid. This is a grid of 6553 points with a mean spacing of 2.1$ \pm 0.1\Deg$. The grid boundary is defined by the equatorial edge of the field of view at Dec. = -16$\Deg$. The red diamond is the location of the Hotspot (\cite{Abbasi:2014lda}), and the green diamond is the location of the energy spectrum anisotropy \cite{Abbasi2018E}. The red line is the supergalactic plane (SGP) and the blue line is the galactic plane (GP).}\label{fig:grid}
\end{figure}

With the drift-diffusion picture of Figure~\ref{fig:deflection} in mind, possible UHECR deflections from grid point ``sources'' were found by a scanned maximization of the significance of energy-angle correlations inside spherical cap sections, or ``wedges,'' using seven years of TA data \cite{LundquistUHECR2019}. This scan was done at each point on an approximately equal 2$\Deg$ spaced grid of 6553 points on the FOV shown in Figure~\ref{fig:grid}~\cite{2006CG.....32.1442T}.

These wedge bins are defined by a maximum angular distance $\delta_j$ from the grid point, $i$, defined by Equation~\ref{eq:dist} and the boundaries of two azimuths defined by Equation~\ref{eq:azimuth} where $B$ is latitude and $L$ is longitude. 

\begin{equation}
\phi_{ij} = \arctan{\frac{\cos{B_i} \sin{(L_i-L_j)}}{\cos{B_j} \sin{B_i}-\sin{B_j} \cos{B_i} \cos{(L_i-L_j)}}}
\label{eq:azimuth}
\end{equation}

The azimuths increase clockwise and a great circle section, or wedge, pointed towards 90$\Deg$ supergalactic latitude (SGB) has an azimuth, $\phi$, of zero. While one pointed towards -90$\Deg$ SGB has a $\phi$ of 180$\Deg$. The azimuthal angle difference, $\Delta\phi_{ij}$, between the wedge pointing direction, $\phi_i$, and the azimuth of an event, $\phi_{ij}$, is shown in Equation~\ref{eq:wedgewidth}. An example wedge is shown in Figure~\ref{fig:wedge1}.

\begin{equation}
\Delta\phi_{ij}=\bmod(|\phi_{ij}-\phi_i| + 180, 360) - 180
\label{eq:wedgewidth}
\end{equation}

This oversampling bin shape means that four parameters must be scanned at every grid point to maximize the pre-trial correlation significance. Even though negative correlations are physically expected by a magnetic field drift-diffusion process; the sign of the correlation, and its strength, are not explicitly scanned for nor restricted. The limits on these parameters are large to account for most conceivable extragalactic magnetic deflection scenarios. The scans are all combinations of the following:
\begin{enumerate}
\item Energy Threshold, $E_i$: 10 to 80~EeV in 5~EeV steps.
\item Wedge Distance, $D_i$$=$$\max(\delta_{ij})$: 15$\Deg$ to 90$\Deg$ in steps of 5$\Deg$.
\item Wedge Direction, $\phi_i$: 0$\Deg$ to 355$\Deg$, 5$\Deg$ steps.
\item Wedge Width, $W_i = 2*\max(|\Delta\phi_{ij}|)$: 10$\Deg$ to 90$\Deg$, 10$\Deg$ steps (5$\Deg$ on each side of $\phi_i$).
\end{enumerate}

Events are inside the wedge if $E_j$$\geq$$E_{i}$ $\&$ $\delta_{ij}$$\leq$$D_i$ $\&$ $-W_i/2 \leq \Delta \phi_{ij} \leq W_i/2$, where $i$ is the index of the grid point. The energy-angle correlation is calculated inside the wedge, $\tau(\delta_{ij},E_j)$, and the parameters ($E_{i},D_{i}, \phi_{i}$, and $W_{i}$) are chosen such that the correlation has the minimum p-value (Equation~\ref{eq:corrsig}). This scan was done using seven years of data. The same bin parameters at each grid point were used for the ten years of data set to test the result.

\subsection{Correlation Example}
\begin{figure*}[htb]
\centering
  \subfigure[]{%
   \includegraphics[width=.5\textwidth]{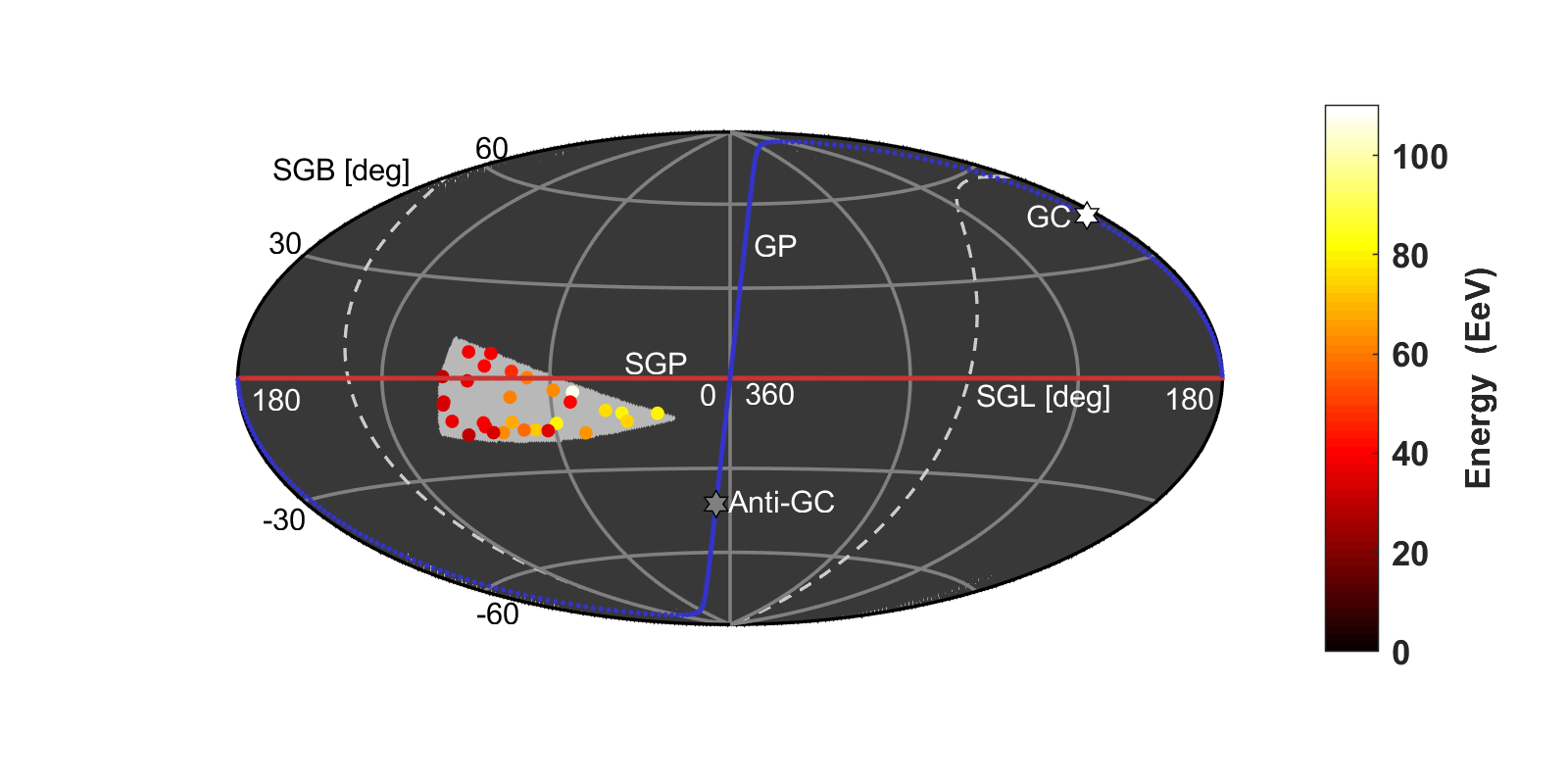}\label{fig:wedge1}}
   \subfigure[]{%
    \includegraphics[width=.49\textwidth]{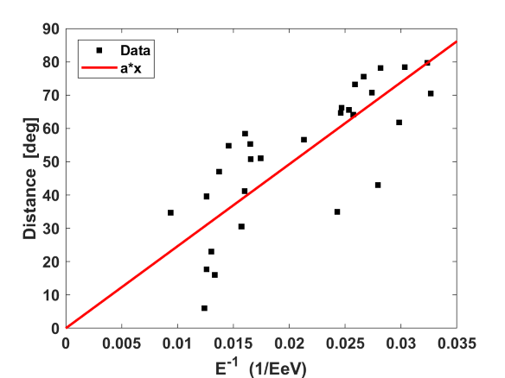}\label{fig:wedgescat1}}
  \caption{(a) A supergalactic Hammer-Aitoff projection of the seven years data spherical cap section, or ``wedge'', with the maximum significance at 18.3$\Deg$ SGB, -12.9$\Deg$ SGL. The correlation strength is $\tau$$=$$-0.675$, and with 29 data events has a pre-trial one-sided significance of 5.5$\sigma$. The energy threshold is $E_i$$\geq$$30$~EeV, wedge width $W_i$$=$$30$$\Deg$, distance $D_i$$=$$80$$\Deg$, and direction $\phi_i$$=$$90$$\Deg$. (b) A scatter plot of $1/E_j$ versus distance $\delta_{ij}$ for events in the wedge. A linear fit (by Equation~\ref{eq:deflect} with $Z$$=$$1$) results in an estimate of $B$$\times$$S=49$~nG*Mpc. If the source is assumed to be the same distance as M82 (3.7~Mpc) with a pure proton emission, then the coherent magnetic field required to cause this deflection would be $B$$=$$13$~nG.}
\end{figure*}

The wedge parameters needed to maximize the correlation significance at each grid point were scanned for using seven years of TA data \cite{LundquistUHECR2019}. For the seven-year data set, the supergalactic coordinates of the most significant correlation of all the grid points is 18.3$\Deg$ SGB (latitude), -12.9$\Deg$ SGL (longitude). This wedge, and the events inside, are shown in Figure~\ref{fig:wedge1}. There are 29 events with energies E$\geq$30~EeV. The pre-trial one-sided significance of a $\tau$$=$$-0.675$ with a sample size of 29 events is 5.5$\sigma$. A scatter plot of energy versus angular distance from the grid point within this wedge is shown in Figure~\ref{fig:wedgescat1}. A linear fit (Equation~\ref{eq:deflect} with $Z$$=$$1$) results in an estimate of $B$$\times$$S=49$~nG*Mpc. If the source is assumed to be at the distance to M82 (3.7~Mpc) with a pure proton emission, the average coherent magnetic field, perpendicular to the FOV, required to cause this deflection would be $B$$=$$13$~nG.

Note however, that the post-trial significance of any single correlation is not expected to be large, as the wedge scan parameter space is large. An individual correlation is not the test of a supergalactic structure.

\section{Simulations} \label{sec:MC}

The same analysis is applied to isotropic simulations in order to calculate the significance of any anisotropy (as described further in Section~\ref{sec:superstruct}). This is a simulation of data with the TA SD configuration while assuming no specific sources or correlation with the supergalactic (or galactic) plane. 

A second simulation is used to demonstrate that the analysis is able to find the hypothesized supergalactic structure. This is a simple toy-model simulation of a supergalactic magnetic sheet that results in an energy-dependent diffusion of events away from the supergalactic plane. This sheet simulation is used to motivate the test statistic that tests the hypothesis of supergalactic sources and magnetic fields; this is further described in Section~\ref{sec:superstruct}. This simulation can also be used to estimate the average coherent field strength between our galaxy and supergalactic sources.

\subsection{Isotropic Simulation}\label{ssec:isoMC}
Each Monte Carlo (MC), and data event, is defined by their energy, zenith angle, azimuthal angle, and trigger time. The latitude and longitude are defined from the center of TA at 39.3$\Deg$ Lat., 112.9$\Deg$ Long. These horizontal coordinates are used to calculate the longitude (SGL) and latitude (SGB) in supergalactic coordinates \cite{Vallado1991}. The MC event sets have a zenith angle distribution of g($\theta$) = sin$(\theta)$cos$(\theta)$ due to the event sampling response of a two-dimensional SD array, a uniform azimuth distribution, and the detection efficiency $\sim$100\% for UHECR E$\geq$$10^{19.0}$~eV. The event trigger times are approximated as a uniform distribution of modified Julian dates from the beginning to the end of the run time due to the approximately $\sim$100\% SD on-time. All of these MC parameter distributions are in good agreement with the data set described in Section \ref{sec:data}.

Detector acceptance, and bias, in the energy spectrum, are taken into account by interpolation sampling of a large set of MC events reconstructed through a surface detector simulation thrown with the average HiRes/TA spectrum \cite{Abbasi:2007sv}. The same cuts applied to the data are applied to these fully simulated events, and there are $\sim$4$\times$10$^{5}$ with energies E$\geq$$10^{19.0}$~eV. The number of events in each isotropic MC event set is the same as data in each 5~EeV bin of the parameter scan of Section~\ref{ssec:wedge}. This simulated data has been shown to reproduce all measured geometric and photoelectric distributions accurately \cite{Ivanov2012}.

The result is that each set of these isotropic MC events simulates the expected data, given the detector configuration and on-time, with no energy anisotropies. These MC sets are used to calculate the post-trial probability of any potential anisotropy signal in the data.

\subsection{Supergalactic Magnetic Sheet}\label{ssec:sheetsim}
\begin{figure}[htb]
\centering
    \includegraphics[width=1\linewidth]{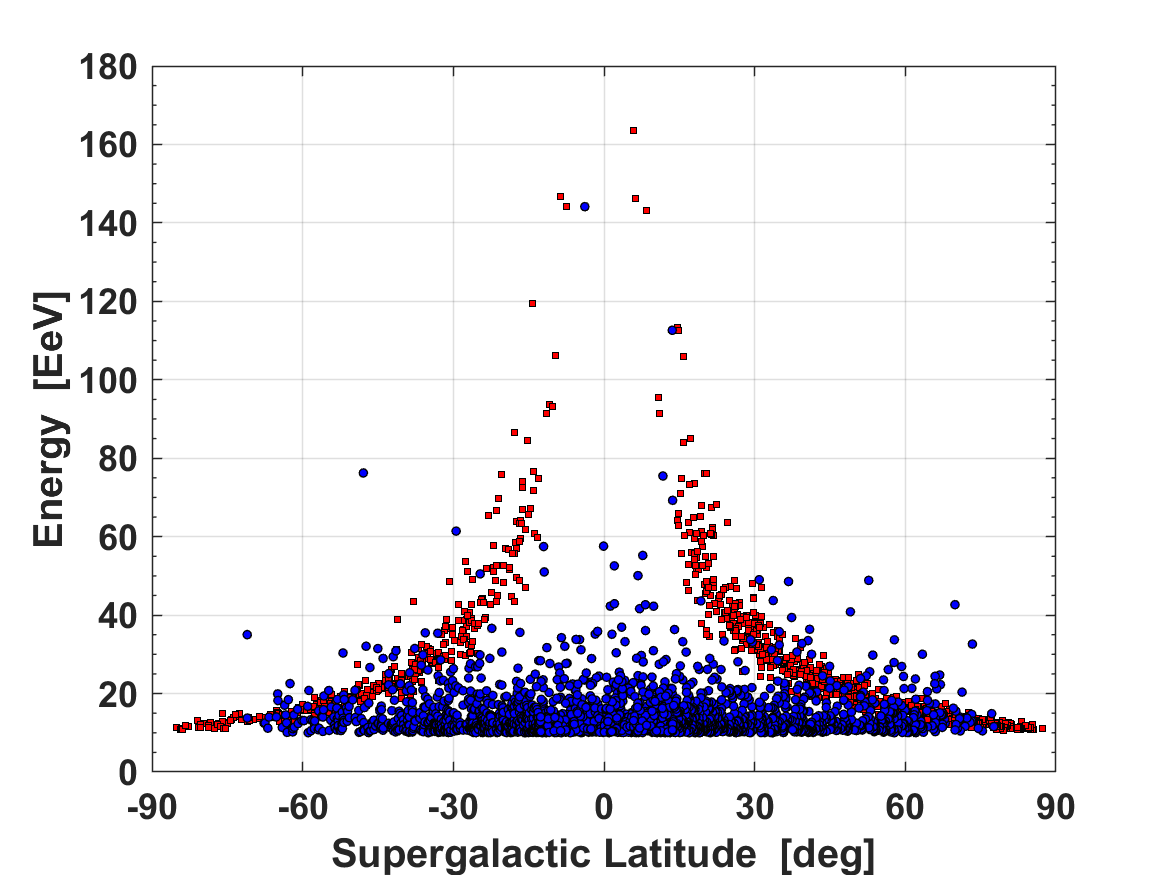}
  \caption{Toy-model supergalactic magnetic sheet simulation. Blue circles are the $F = 65.7\%$ isotropic fraction of MC events. Red squares are the anisotropic MC events magnetically diffused away from the supergalactic plane with $B$$\times$$S$$=$$18.47$~nG*Mpc. Overall, event positions are isotropic, and the energy spectrum is created according to the published HiRes/TA result.}\label{fig:supersim}
\end{figure}

A simple toy-model simulation of an intervening supergalactic magnetic sheet, between the galaxy and UHECR sources, is made by taking the isotropic event sets of Section~\ref{ssec:isoMC} and embedding event deflections (assigning distance correlated energies) in supergalactic latitude (SGB), proportional to $1/energy$, for a fraction of events. The coordinates of the MC events are isotropic and unchanged in the procedure. The approximate apparent deflection from the source of a charged particle in a coherent magnetic field is from Equation~\ref{eq:deflect}. An example of the resulting simulation is shown in Figure~\ref{fig:supersim}. 

The event deflections, $\delta_B$, from supergalactic latitude $SGB=0\Deg$ are calculated for each MC event energy in the set, assuming a proton composition ($Z$$=$$1$) and a particular $B$$\times$$S$, according to Equation~\ref{eq:deflect}. Additionally, some random field noise was added by smearing the $\delta_B$ with a 5$\Deg$ standard deviation Gaussian. Then each event is assigned an energy based on its angular distance from the supergalactic plane ($min$[$\delta_B$-$\delta_{SGB=0}$]). The beginning, and final, simulation is isotropic with respect to the supergalactic longitude (SGL).

After the assignment of an energy to each event position, those with an assigned position-deflection error greater than 10$\Deg$ are added to the isotropic proportion. This threshold adds additional random field noise in the simulation. This cut also results in a harder spectrum for the deflected events (red event in Figure~\ref{fig:supersim}) i.e. higher energy events on average closer to the supergalactic plane. This supergalactic energy bias is due to the lower number of high energy events resulting in a better fit to a supergalactic magnetic deflection at higher energies (due to the boundary conditions of the energy spectrum and position isotropy).

A supergalactic sheet simulation, with an $F$$=$$65.7\%$ isotropic fraction and $B$$\times$$S$$=$$18.47$~nG*Mpc, is shown in Figure~\ref{fig:supersim}. These parameters are the result of selecting a random MC that looks similar to the data result and the choice of a proton composition. Note again that this is only an anisotropy of energies in supergalactic latitude as event positions are isotropic, and the total energy spectrum is unchanged.

The intent of this toy-model simulation is to show that the analysis method is sensitive to an energy symmetry caused by some kind of magnetic deflection structure correlated with the supergalactic plane. It is not intended to reproduce all aspects of actual data.

\section{Supergalactic Structure} \label{sec:superstruct}

No single correlation tests the hypothesis that sources and magnetic fields have a relation to local large scale structure. And no single correlation can be significant considering the average $\sim$60,000 scan parameter combinations at all 6553 grid points. 

\begin{figure}[htb]
\centering
  \subfigure[]{%
   \includegraphics[width=.5\textwidth]{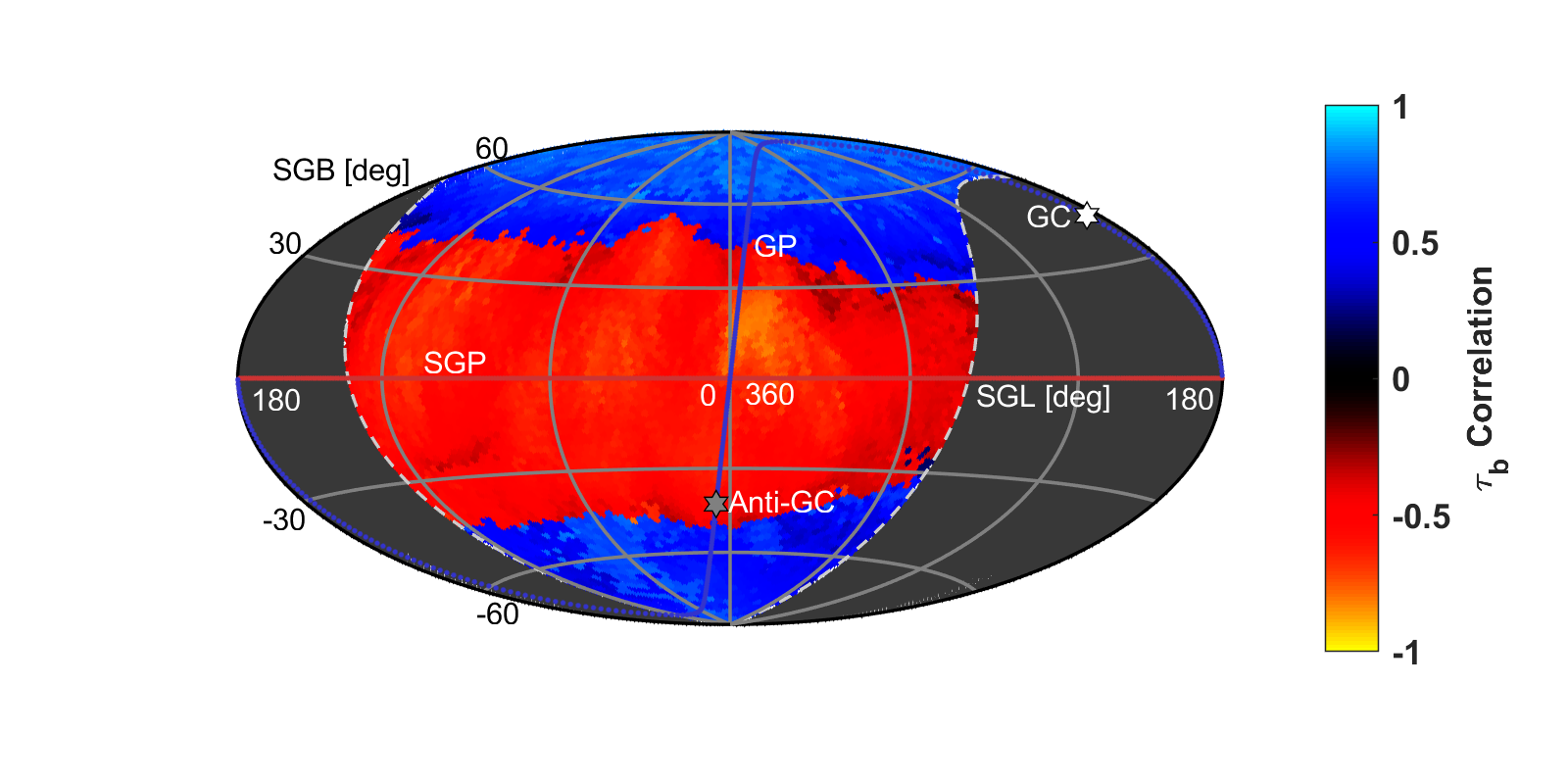}\label{fig:corrtauMC}}
   \subfigure[]{%
    \includegraphics[width=.5\textwidth]{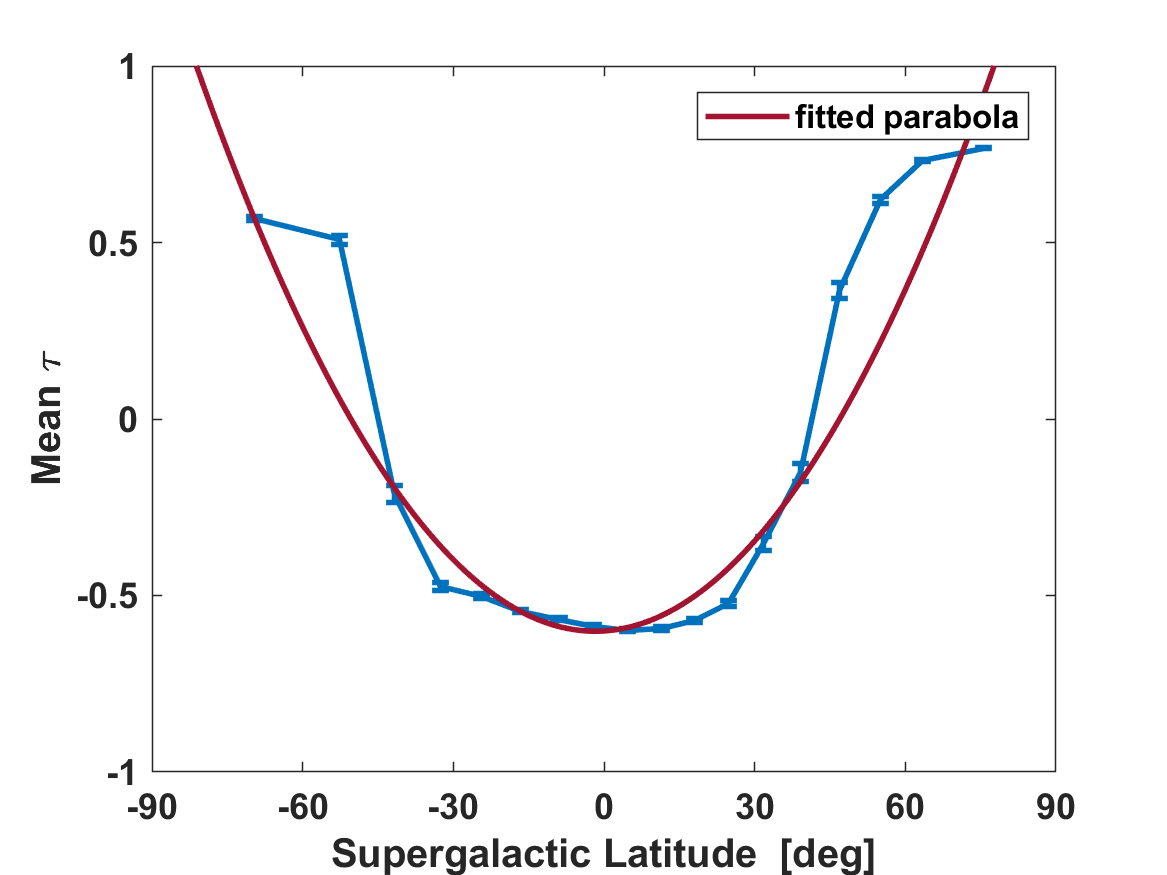}\label{fig:MCave}}
  \caption{A supergalactic magnetic sheet simulation. (a) Projection of the correlation strength $\tau$ for all grid points. Solid curves indicate the galactic plane (GP) in blue and supergalactic plane (SGP) in red. (b) Mean $\tau$ inside equal solid angle bins of supergalactic latitude (SGB). The parabolic fit ($y = a (x-x_0)^2 + y_0$) shows the curvature parameter, $a$, chosen as the test statistic. $a$$=$$2.5$$\times$$10^{-4}$.}
\end{figure}

As an example of what can be expected, large scale behavior is demonstrated by the oversampled wedge correlation result for the supergalactic sheet simulation shown in Figure~\ref{fig:corrtauMC} ($F$$=$$65.7\%$ isotropy and $B$$\times$$S$$=$$18.47$~nG*Mpc). 

It can be seen via this simple model in the projection of $\tau$ that if there are magnetically induced energy-angle correlations clustered in the supergalactic plane, negative correlation wedges will be close to the supergalactic plane. Furthermore, since negative correlations viewed from the opposite direction appear as positive correlations (as can been seen by Equation \ref{eq:kendall} for $[x$$=$E, y$=$D$] \rightarrow [$x$=$E, y$=$-D$])$, positive correlations are expected at large distances from the supergalactic plane.

\subsection{Significance Test}

Though a test for a supergalactic structure of energy-angle correlations is not necessarily $a$ $priori$ obvious, the supergalactic sheet toy-model leads to a reasonable answer. The mean $<$$\tau$$>$ inside equal solid angle bins of angular distance ($SGB_i$) from the supergalactic plane (SGP) shows that three features are relevant for the supergalactic hypothesis - the minimum average $\tau$, the minimum location being near the SGP, and the symmetry of $\tau$ around the SGP. Using all three features to calculate the data significance would be overfitting the problem. One test statistic is preferable though it should be correlated with these three supergalactic structure features. The single parameter chosen to test the supergalactic structure hypothesis is the curvature parameter, ``$a$,'' of a parabolic fit (y = $a (x-x_0)^2 + y_0$) to the $<$$\tau$$>$. 

\begin{figure}[htb]
\centering
  \subfigure[]{%
   \includegraphics[width=.5\textwidth]{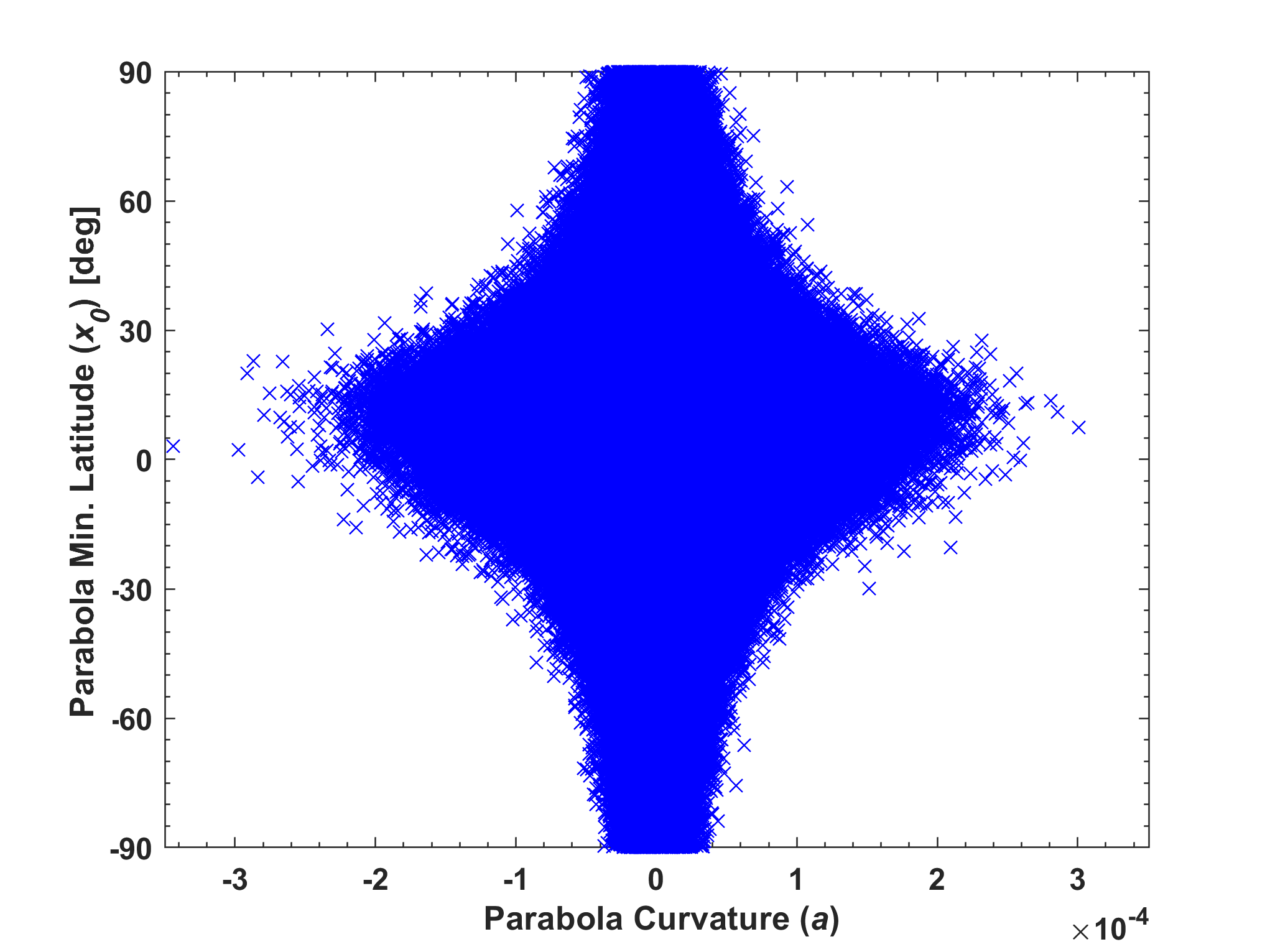}\label{fig:curvemin}}
   \subfigure[]{%
    \includegraphics[width=.5\textwidth]{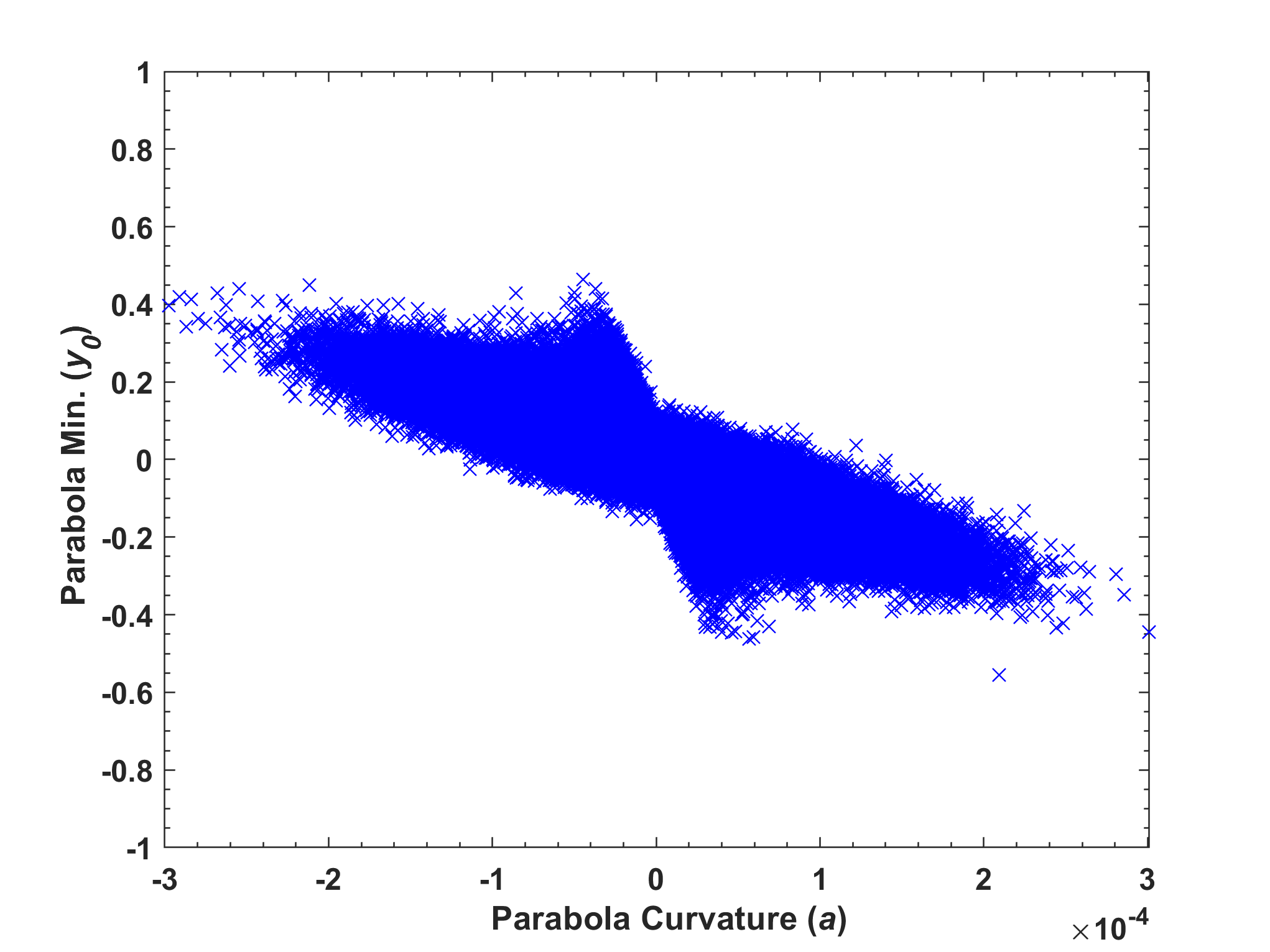}\label{fig:curveminval}}
  \caption{The behavior of the three mean $\tau$ parabola fit parameters (Figure~\ref{fig:MCave}) with respect to each other in random isotropic MC simulations. (a) The parabola fit curvature, $a$, versus minimum supergalactic latitude (SGB), $x_0$, shows that a high curvature tends to a minimum near the supergalactic plane. (b) The parabola fit curvature, $a$, versus the fit minimum value, $y_0$, shows that a high curvature tends to a higher magnitude negative mean $\tau$.}
\end{figure}

The curvature, ``$a$'', is simply the lowest order Taylor expansion term that can describe the symmetry around the SGP shown in simulation (Figure~\ref{fig:MCave}). Due to the boundaries of $|\tau|$$\leq$1 and $|SGB|$$<$90$\Deg$, greater correlation curvature, $a$, corresponds to a minimum, $x_0$, closer to the supergalactic plane, as shown in Figure~\ref{fig:curvemin}. A larger curvature $a$ also means that the minimum negative correlation averages are greater in magnitude, $y_0$, as shown in Figure~\ref{fig:curveminval}. The parabola minimum $y_0$ has no correlation with the minimum supergalactic latitude (SGB). These relationships justify the use of the parabola curvature ``a'' as the single test statistic for a conservative estimate of the significance of supergalactic energy-angle correlations.

The fit on the large scale behavior of the correlation strength, $\tau$, is used because it is not explicitly scanned for and contains more information by its sign ($\pm$) than the pre-trial significance. The pre-trial significance of the correlations is not used in this analysis so that the significance test is independent of the wedge scan for the maximum significance of individual energy-angle correlations. 

To calculate the data significance of a supergalactic structure of energy-angle correlations, the analysis described above was applied to the data and the isotropic MC sets. The number of MC sets with a correlation curvature $a$ greater than the data gives the probability of the measured supergalactic structure of energy-angle correlations if there actually isn't such structure i.e. if it is a statistical fluctuation in the data.

\section{Data Set} \label{sec:data}
For this analysis, Surface Detector (SD) data recorded between May 11 of 2008 and 2019 is used. Data from 2016 is excluded due to issues with SD communication towers that created a significant day to day change of the trigger delay variance within each day of the year. This introduced non-physical equatorial anisotropies that are non-trivial to compensate for. 

The reconstruction method used for these events is the same as the ``Hotspot'' and energy spectrum anisotropy results (\cite{Abbasi:2014lda}, \cite{Abbasi2018E}). The energy of reconstructed events is determined by the SD array and renormalized by 1/1.27 to match the calorimetrically determined fluorescence detector energy scale (\cite{AbuZayyad:2012ru}).

Due to the inclusion of lower energy events, down to 10$^{19.0}$~eV, tighter data cuts than the hotspot analysis are required for good zenith angle and energy resolutions. After cuts, there were 3018 events in the seven-year data set, and there is a total of 4321 events using ten years of data. Events in the data set match the following criteria:
\begin{enumerate}
\itemsep-.3em
\item E$\geq$$10^{19.0}$~eV (where detection efficiency is $\sim$100\%).
\item At least four SDs triggered.
\item Zenith angle of arrival direction $<$55$\Deg$.
\item Shower lateral distribution fit $\chi^2/dof$$<$10. 
\item Reconstructed pointing direction error $<$5$\Deg$.
\item Shower core $>$1.2 km from array boundary.
\end{enumerate}

The additional cuts on pointing direction error and boundary distance improve the agreement between the distribution of zenith angles and the geometrical zenith angle distribution g($\theta$) = sin$(\theta)$cos$(\theta)$. The azimuthal angle distribution is in very good agreement with the theoretical uniform distribution. The geometrical zenith angle distribution is due to the flat detector plane, the uniform azimuthal angle distribution, and the detection efficiency $\sim$100\% for UHECR with energies E$\geq$$10^{19.0}$~eV. The energy spectrum is also in good agreement with the published spectrum (\cite{AbuZayyad:2012ru},\cite{TelescopeArray:2014zca}). And finally, the event trigger times are in good agreement with the uniform time distribution used for the isotropic MC of Section \ref{ssec:isoMC}.

The energy resolution and pointing direction resolution of events in the data set range from $\sim$10 to 15\% and $\sim$1.0$\Deg$ to 1.5$\Deg$, respectively, depending on core distance from the array boundary and improve with increasing energy. These resolutions are sufficient to search for large-scale and intermediate-scale UHECR energy anisotropies.
 
\section{Results}\label{sec:result}

\begin{figure}[h]
\centering
  \subfigure[]{%
   \includegraphics[width=.5\textwidth]{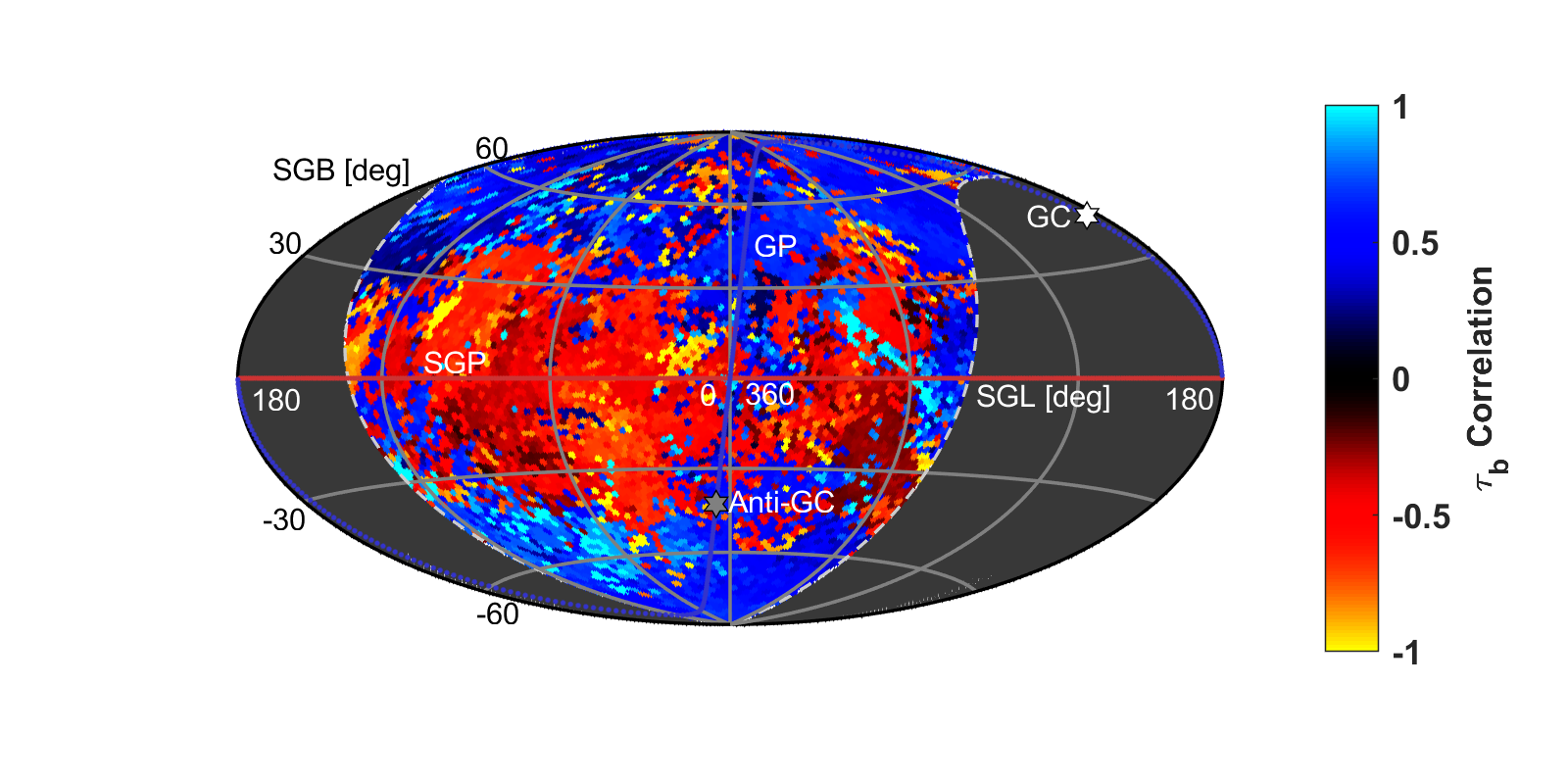}\label{fig:corrtau}}
   \subfigure[]{%
    \includegraphics[width=.5\textwidth]{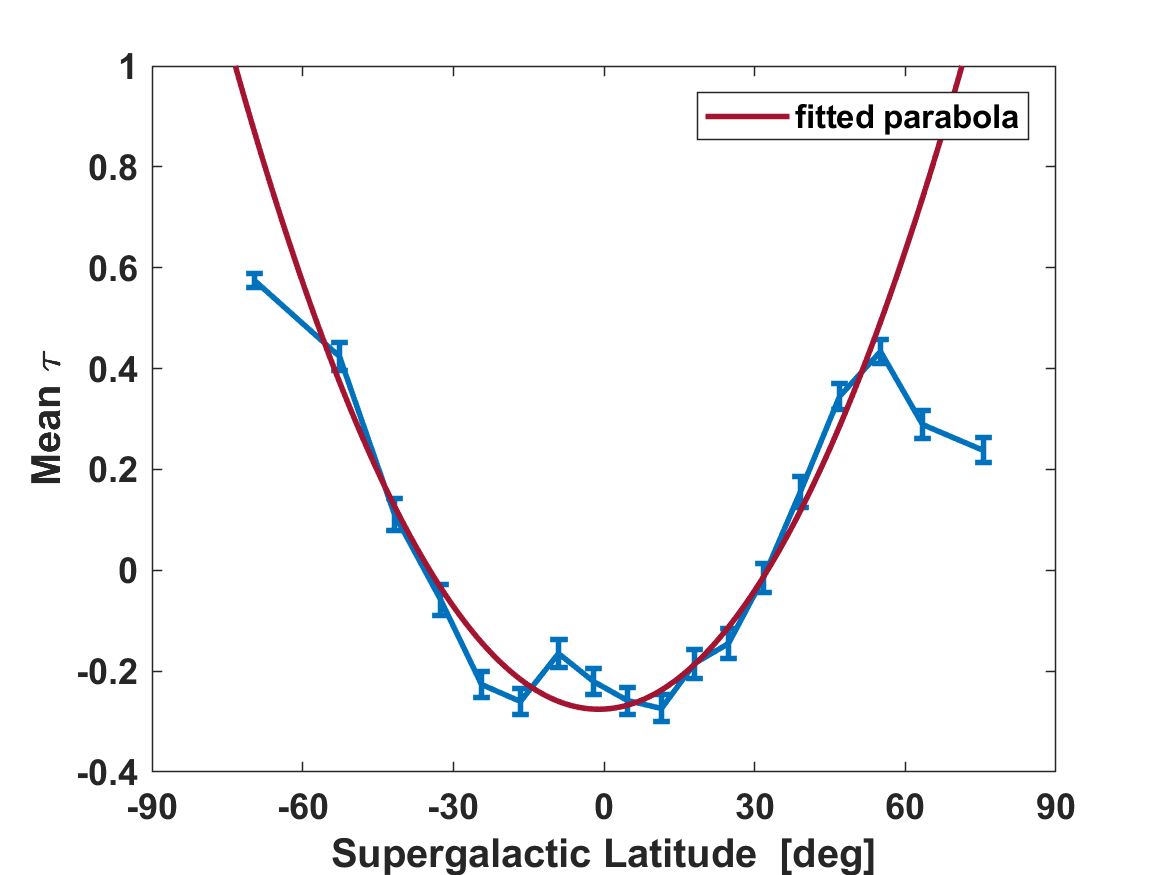}\label{fig:dave}}
  \caption{Seven year-data result. (a) Projection of the correlation strength $\tau$ for all grid points. Negative correlations expected for magnetic deflections are apparent around the supergalactic plane. Solid curves indicate the galactic plane (GP) in blue and supergalactic plane (SGP) in red. White and grey hexagrams indicate the galactic center (GC) and anti-galactic center (Anti-GC) respectively. (b) Mean $\tau$ inside equal solid angle bins of supergalactic latitude (SGB). The correlation curvature is $a$$=$$($$2.45$$\pm 0.15)\times$$10^{-4}$.}
\end{figure}

\begin{figure}[h]
\centering
  \subfigure[]{%
   \includegraphics[width=.5\textwidth]{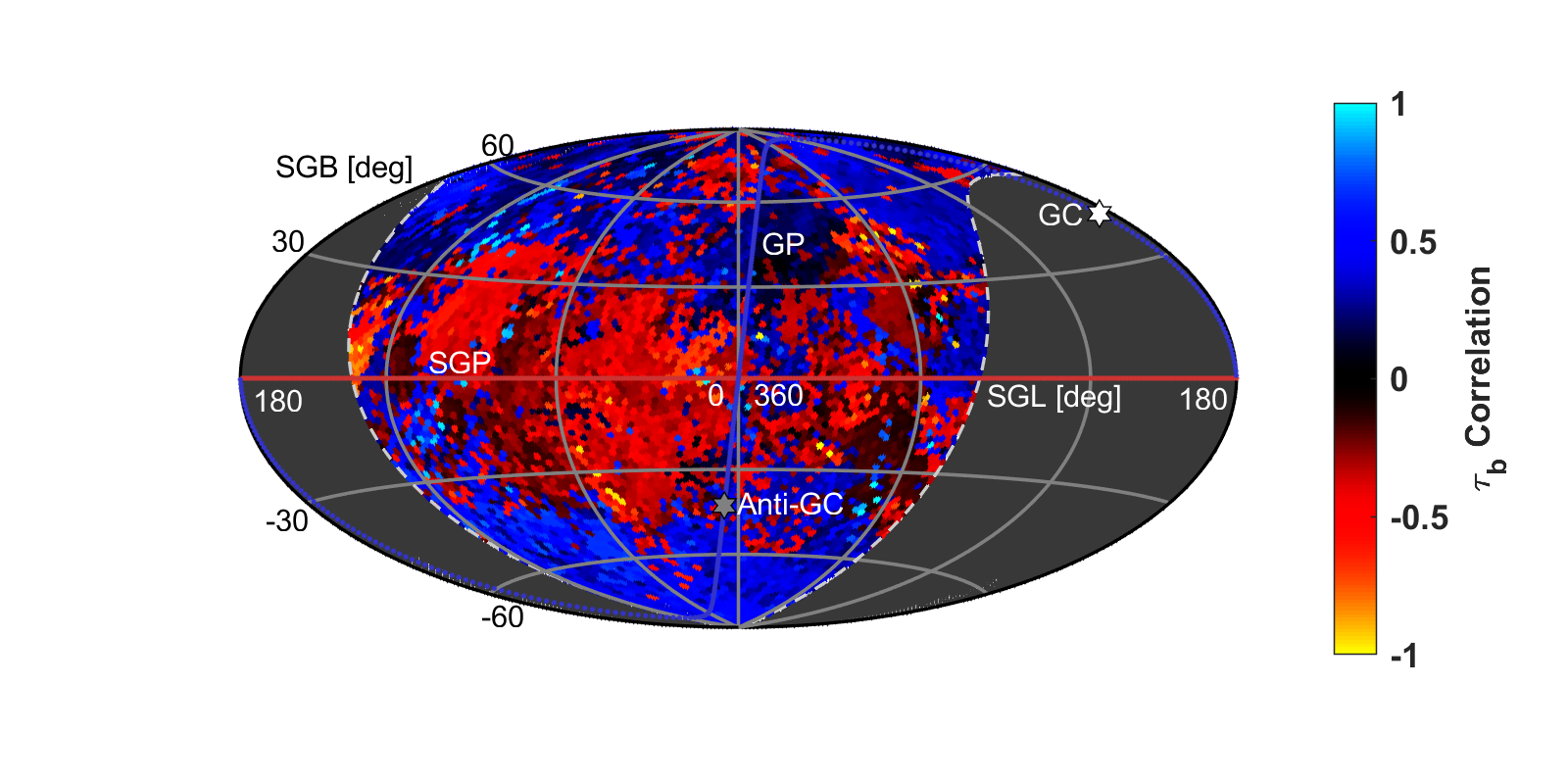}\label{fig:corrtau10}}
   \subfigure[]{%
    \includegraphics[width=.5\textwidth]{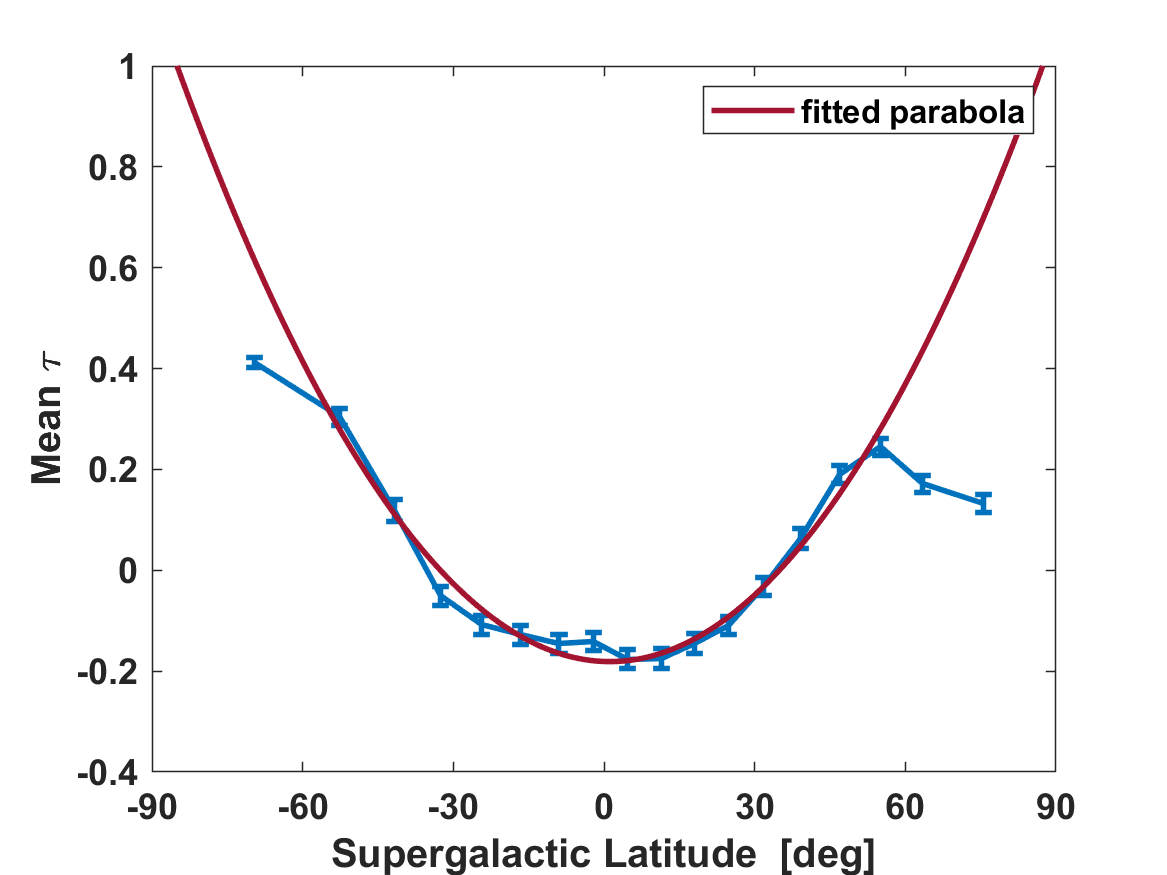}\label{fig:dave10}}
  \caption{Ten years of data result. (a) Projection of the correlation strength $\tau$ for all grid points. Negative correlations expected for magnetic deflections are apparent around the supergalactic plane. (b) Mean $\tau$ inside equal solid angle bins of supergalactic latitude (SGB). The correlation curvature is $a$$=$$($$1.60$$\pm 0.09)\times$$10^{-4}$.}
\end{figure}

The resulting data energy-angle correlations for seven years of data are shown in Figure~\ref{fig:corrtau} and ten years of data is shown in Figure~\ref{fig:corrtau10}. Individual correlations with the highest pre-trial significance are negative, which means that there is a trend for the angular distance to increase with decreasing energy. This trend is the expectation for a grid point that happens to be near a source of magnetically scattered UHECR events. It can be seen that the negative $\tau$ correlations appear themselves well correlated with the supergalactic plane.

Figure~\ref{fig:dave} shows the seven-year data result of the mean $\tau$ correlation inside equal solid angle bins parallel to the supergalactic plane (SGP). The parabolic fit curvature is $a$$=$$($$2.45$$\pm 0.15)\times$$10^{-4}$ with a minimum at $-0.5\Deg$ SGB. According to the $R^2$ (Coefficient of Determination) goodness-of-fit the model predicts 88$\%$ of the variance of the data. The data correlations have a very similar form to that of the supergalactic magnetic sheet simulation, shown in Figure~\ref{fig:corrtauMC}, that has $a$$=$$2.5$$\times$$10^{-4}$ with a minimum at $-1.7\Deg$ SGB.

Previously, by applying this analysis to isotropic MC sets (using data positions and random energies) the number of MC with an $a$ parameter greater than data was two out of 200,000 trials which resulted in a post-trial significance of the supergalactic structure of multiplets of $\sim$4$\sigma$ \cite{LundquistUHECR2019}.

Figure~\ref{fig:dave10} shows the ten years of data mean $\tau$ correlation with no new scan of wedge parameters for maximum correlation significances. The parabola curvature is $a$$=$$($$1.60$$\pm 0.09)\times$$10^{-4}$, and the minimum is at $1.1\Deg$ SGB. According to the $R^2=0.91$ goodness-of-fit the model predicts 91$\%$ of the variance of the data. It can be seen that the correlations are similar to the seven-year result, though the supergalactic structure may not be quite as significant.

\subsection{Significance of Supergalactic Structure}\label{sec:sigmethod}
\begin{figure}[htb]
\centering
    \includegraphics[width=1\linewidth]{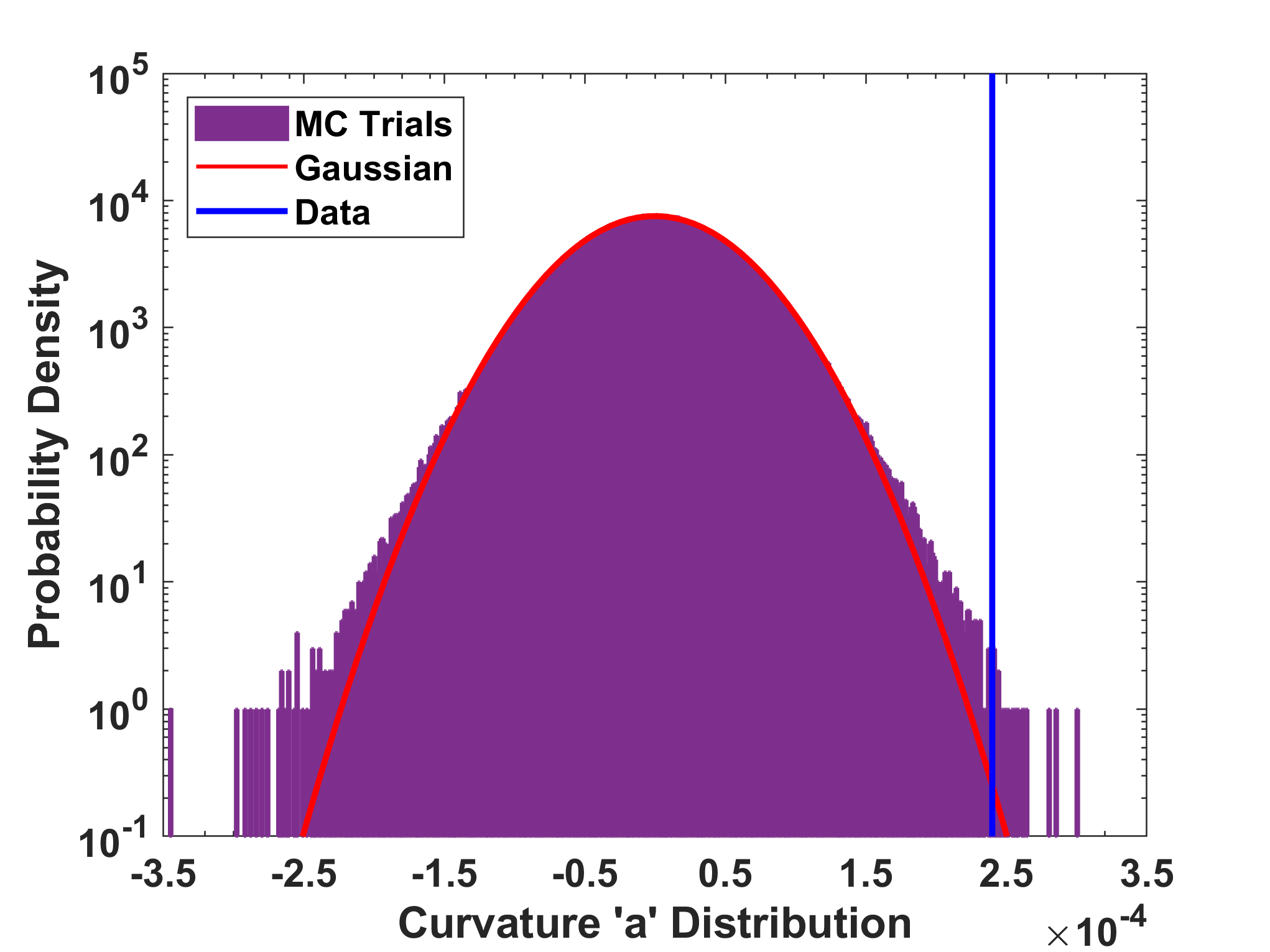}
  \caption{The distribution of the curvature parameter $a$ of the mean $\tau$ parabola chosen as the supergalactic structure of multiplets test statistic for 1,000,000 isotropic MC sets. The purple bars are the MC probability distribution function (PDF). The red line is a Gaussian distribution fit to the MC distribution. The curvature for the data is $a$$=$$2.45$$\times$$10^{-4}$ shown as a blue vertical line. There are 14 MC with a larger curvature than data, which gives a significance of 4.2$\sigma$.}\label{fig:adist}
\end{figure}

By applying this analysis to isotropic MC sets, as described in Section~\ref{ssec:isoMC}, and counting the number of MC with an $a$ parameter larger than data (Figure~\ref{fig:dave}), the post-trial significance of the supergalactic structure of multiplets can be found. The resulting $a$ distribution of 1,000,000 MC sets is shown in Figure~\ref{fig:adist} for the seven-year data statistics and energy-angle correlation significance scan.

For the seven-year data analysis there are 14 MC sets with a larger curvature than data, which results in the significance of a supergalactic structure of multiplets of $\sim$4.2$\sigma$.

\begin{figure}[htb]
\centering
    \includegraphics[width=1\linewidth]{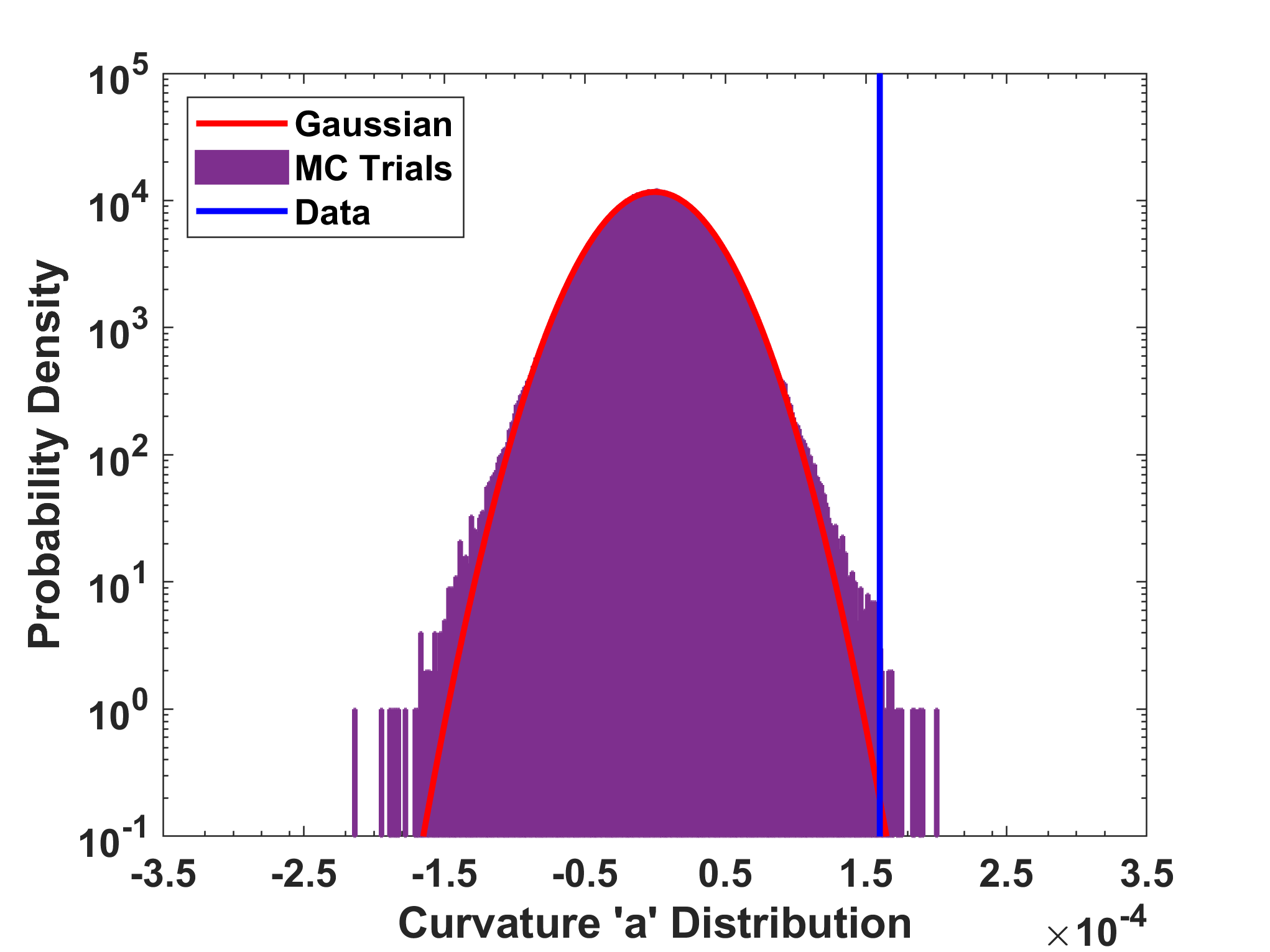}
  \caption{The distribution of the curvature parameter $a$ of the mean $\tau$ parabola chosen as the supergalactic structure of multiplets test statistic for 900,000 isotropic MC sets. The purple bars are the MC PDF. The red line is a Gaussian distribution fit to the MC distribution. The curvature for the data is $a$$=$$1.60$$\times$$10^{-4}$ shown as a blue vertical line. There are 22 MC with a larger curvature than data, which gives a significance of 4.1$\sigma$.}\label{fig:adist2}
\end{figure}

For the ten years of data with no updated wedge correlation significance scan, the resulting $a$ distribution of 1,000,000 MC sets is shown in Figure~\ref{fig:adist2}. The distribution has a smaller standard deviation due to no new scan for energy-angle correlation significances. The result is smaller $\tau$ on average. There are 22 MC sets with a larger curvature than data, shown in Figure~\ref{fig:dave10}, which results in the significance of a supergalactic structure of multiplets of $\sim$4.1$\sigma$.

The total number of MC sets that were used to calculate the significance was limited by the computing time necessary for each simulation. Overall, the number of correlations calculated was 4$\times$10$^{14}$ and this took more than 200 years of equivalent CPU computing time.

\subsection{Scan Parameter Distributions}\label{sec:params}
\begin{figure}[h]
\centering
  \subfigure[]{%
   \includegraphics[width=1\linewidth]{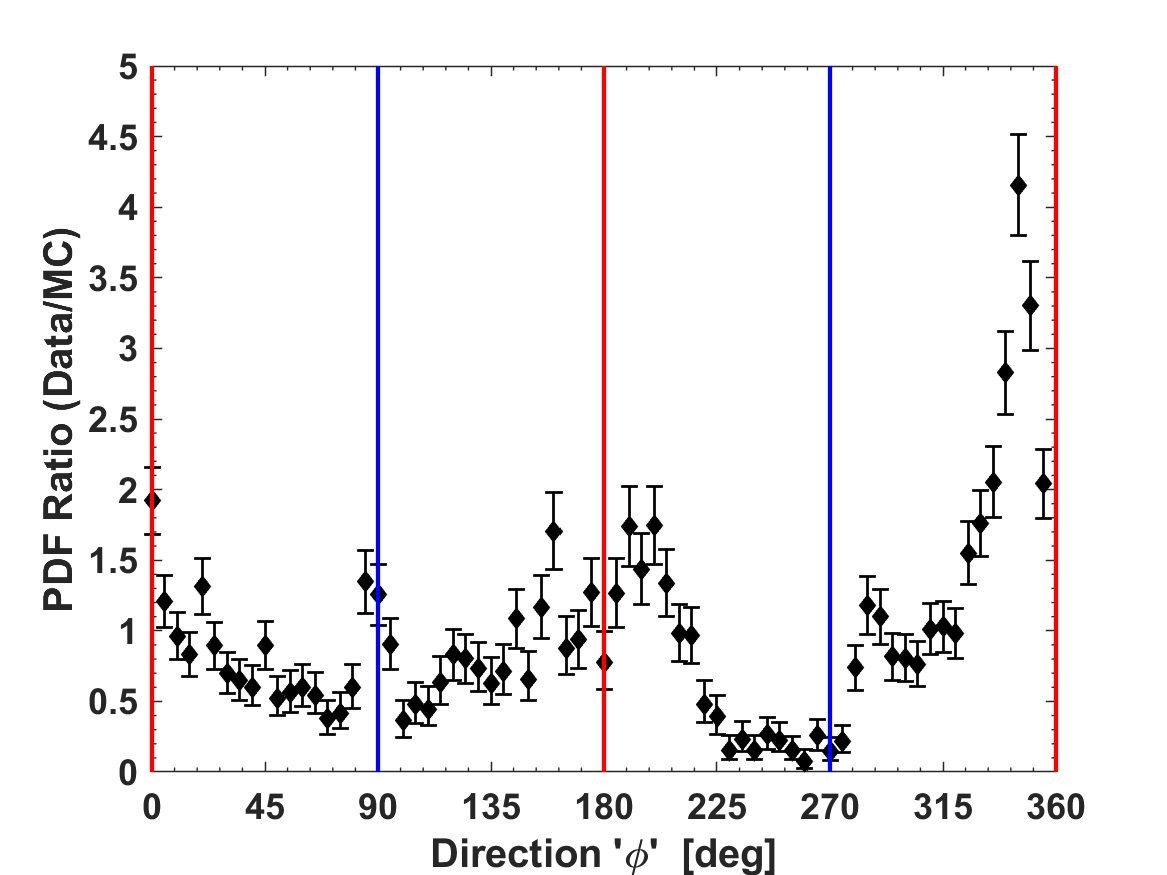}\label{fig:direction}}
   \subfigure[]{%
    \includegraphics[width=1\linewidth]{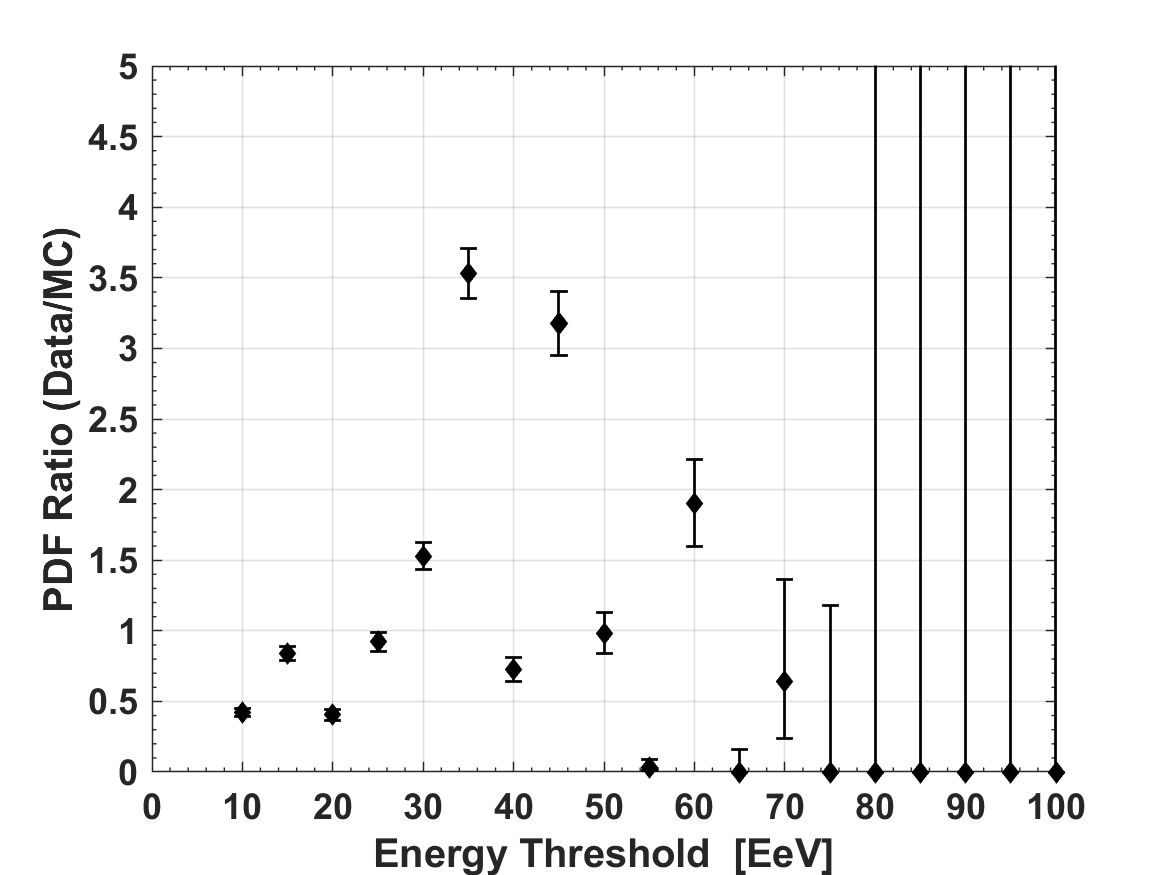}\label{fig:energy}}
  \caption{PDF ratio plot of scanned parameters. (a) Wedge pointing direction parameter, $\phi$. This distribution provides further indication of the supergalactic structure of multiplets. The blue vertical lines are parallel to the SGP. The red lines are perpendicular to the SGP. Two significant peaks can be seen nearly perpendicular, and two smaller peaks near parallel, to the SGP. (b) Energy threshold, $E$. The three largest deviations are at 35~EeV, 45~EeV, and 60~EeV. This distribution may provide information regarding UHECR sources and intervening fields.}\label{fig:ratios}
\end{figure}

Clues about UHECR sources, and intervening fields, may be found from the maximum significance wedge scan parameters of the apparent magnetic deflection multiplets. Due to the significance maximization, there is a bias towards greater statistics, as can be seen in Equation~\ref{eq:corrsig}, so the data is compared to isotropic MC by taking the ratio of the parameter probability distribution functions (PDF) (normalized histograms of data divided by MC). The PDF ratio shows how many times more likely a scan parameter value is to be found in data than isotropic MC. PDF ratio plots for wedge pointing direction and energy threshold parameters are shown in Figure~\ref{fig:ratios}.

These ratios are done for negative energy-angle correlations at grid point positions $\lvert SGB\rvert\leq40\Deg$ (about the boundary where the average correlation is zero as shown in Figure \ref{fig:corrtau10}) and have a linear fit to $1/E$ versus angular distance with an $R^2$$>$$0$ (Figure~\ref{fig:wedgescat1}). An $R^2$$>$$0$ is a better fit than a horizontal line, and the $\delta$$\propto$$1/E$ model explains some of the variance of the data inside the wedge. For data, there are 2045 correlations used and greater than 3.99$\times$$10^{8}$ for MC.

The data distribution of wedge pointing directions, Figure~\ref{fig:direction}, provides further indication of a supergalactic structure with four deviations seemingly correlated with the supergalactic plane (SGP). Two larger peaks are approximately perpendicular to the SGP ($\sim$195$\Deg$ and $\sim$345$\Deg$), and two smaller peaks close to parallel ($\sim$90$\Deg$ and $\sim$285$\Deg$). These peaks suggest an overall diffusion of low energy events away from the supergalactic plane, similar to the supergalactic magnetic sheet simulation of Section~\ref{ssec:sheetsim}.

The data distribution of the energy threshold parameters may provide information regarding UHECR sources and intervening fields. The median energy threshold is 30~EeV, and the three largest deviations from the isotropic distribution are at 35~EeV, 45~EeV, and 60~EeV. The 60~EeV peak appears to correspond to the 57~EeV threshold of the TA hotspot analysis \cite{Abbasi:2014lda}.

The median energy threshold of 30~EeV is above the significant Pierre Auger Observatory (PAO) large scale dipole measurements in \cite{Aab_2018} at 8 EeV, which is consistent with the localized intermediate-scale energy-angle magnetic deflections in this analysis. 

The 39~EeV cutoff for maximum event correlation with starburst galaxies, reported by the PAO in \cite{Aab:2018chp}, may be related to the 35~EeV and 45~EeV peaks. 

These threshold deviations from isotropy are also consistent with the result using AGASA data that showed a possible large scale cross-correlation between UHECR and the supergalactic plane between 50 and 80~EeV energy bins \cite{Burgett:2003yg}. Adjusting the AGASA energy scale to the TA energy scale by multiplying by 0.75, this becomes 38 and 60~EeV energy bins. 

The data distributions of wedge angular distance, $D$, and width, $W$, do not show any significant deviations from isotropy. 

\subsection{M82 Galaxy as Anisotropy Source}\label{sec:M82}

\begin{figure}[htb]
\centering
  \subfigure[]{%
   \includegraphics[width=.5\textwidth]{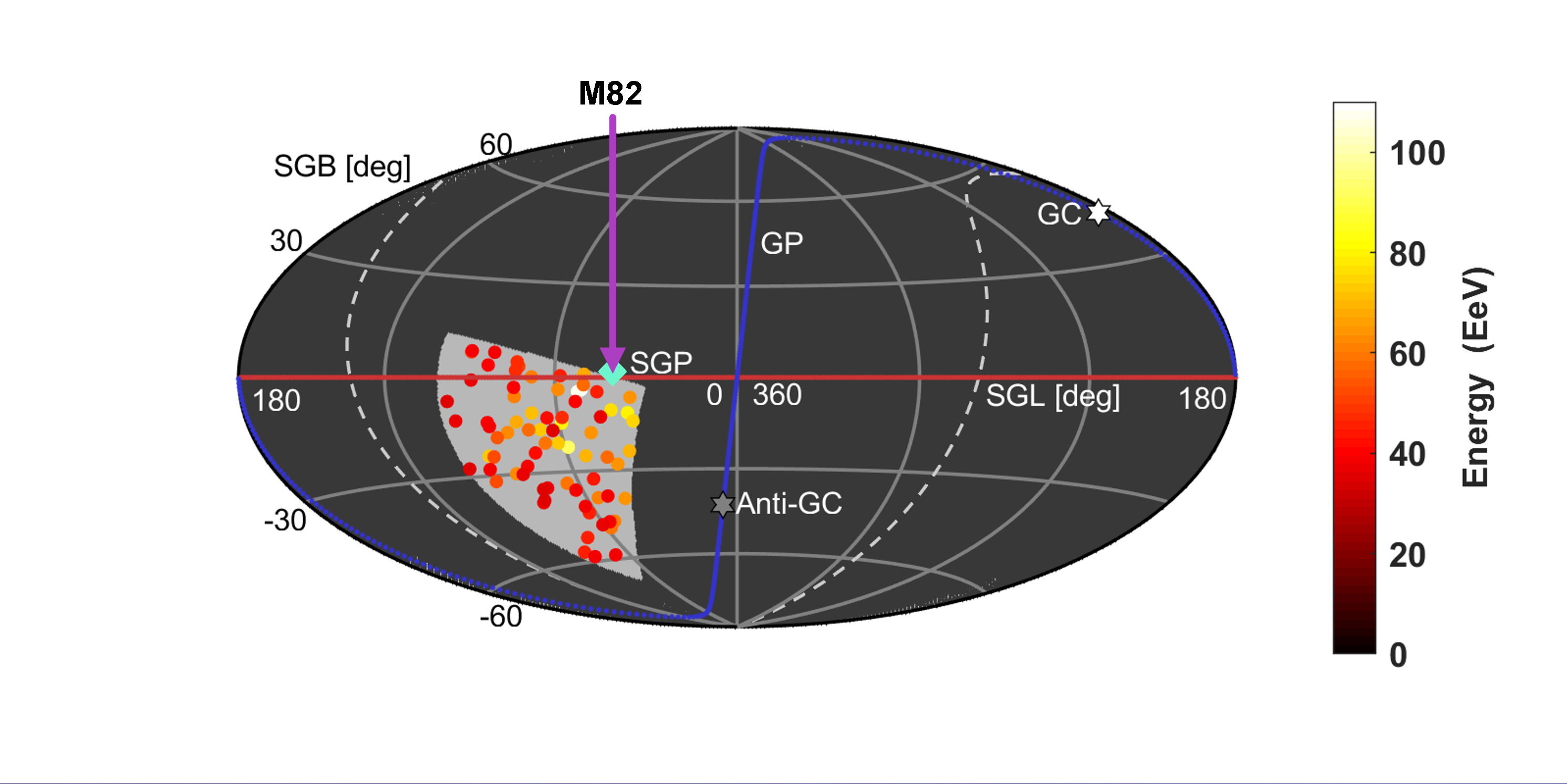}\label{fig:wedge2}}
   \subfigure[]{%
    \includegraphics[width=.5\textwidth]{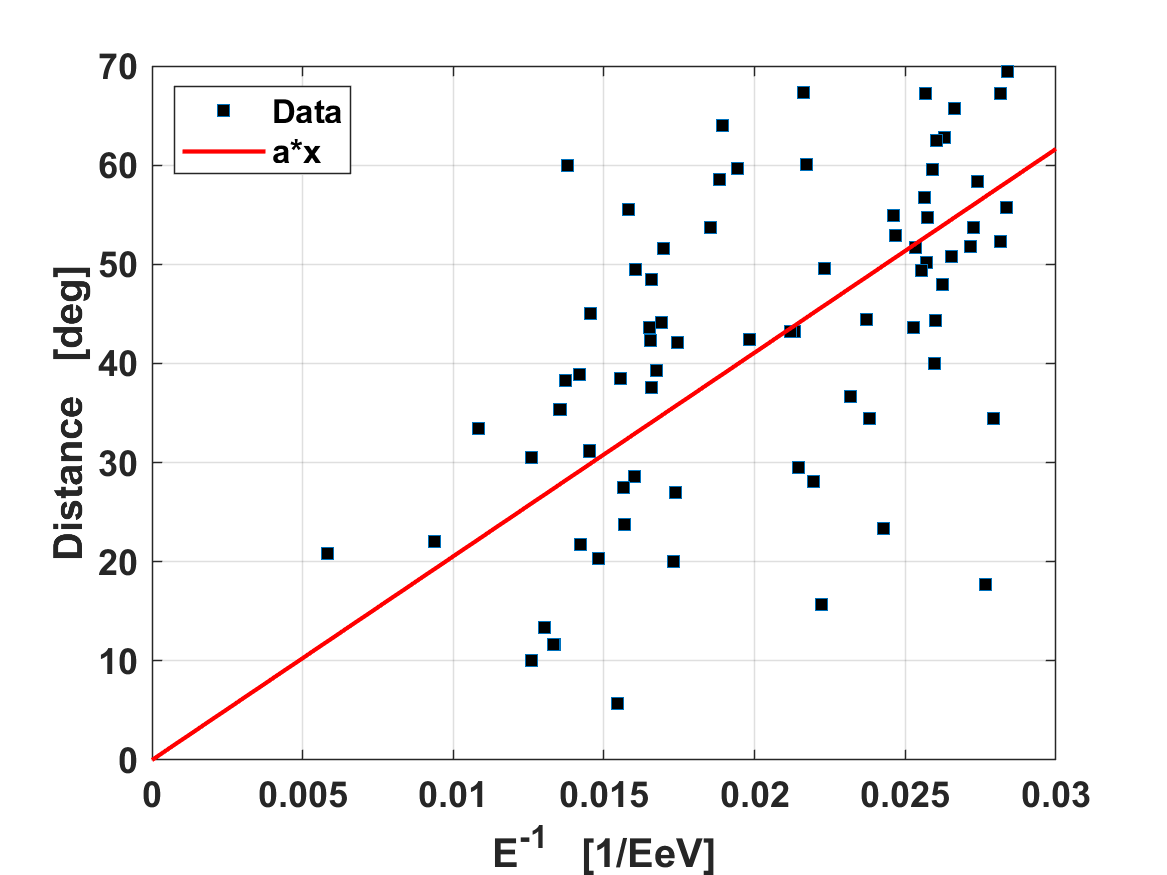}\label{fig:wedgescat2}}
  \caption{(a) Supergalactic projection of the ten years data most significant ``wedge'' multiplet at 30.3$\Deg$ SGB, -3.2$\Deg$ SGL. The correlation $\tau$$=$$-0.412$ with 75 data events has a pre-trial one-sided significance of 5.10$\sigma$. This significance is an increase from 4.58$\sigma$ at this grid point, with seven years of data. The energy threshold is $E_i$$\geq$$35$~EeV, wedge width $W_i$$=$$90$$\Deg$, angular distance $D_i$$=$$70$$\Deg$, and direction $\phi_i$$=$$120$$\Deg$. The blue diamond is the location of the starburst galaxy M82. (b) Scatter plot of $1/E_j$ versus angular distance $\delta_j$ in the wedge. A linear fit (by Equation~\ref{eq:deflect} with $Z$$=$$1$) results in an estimate of $B$$\times$$S=41$~nG*Mpc. If the source is assumed to be at the same distance to M82 (3.7~Mpc) with a pure proton emission, then the average coherent magnetic field required to cause this deflection would be $B$$=$$11$~nG.}
\end{figure}

The most significant single correlation using ten years of SD data is at 30.3$\Deg$ SGB, -3.2$\Deg$ SGL, and shown in Figure~\ref{fig:wedge2}. With 75 events (E$\geq$35~EeV) and $\tau$$=$$-0.412$, it has a pre-trial significance of 5.10$\sigma$. This significance is an increase from 4.58$\sigma$ at this grid point, with seven years of data using the same wedge and energy threshold parameters. Figure~\ref{fig:wedgescat2} shows a scatter plot of energy versus angular distance. A linear fit (Equation~\ref{eq:deflect} with $Z$$=$$1$) results in an estimate of $B$$\times$$S=41$~nG*Mpc. 

Recently, the Pierre Auger Observatory (PAO) has stated that the likeliest source of events with E$>$39~EeV are starburst galaxies \cite{Aab:2018chp}. The most significant correlation reported here is 11.3$\Deg$ from M82 (as shown by the blue diamond in Figure \ref{fig:wedge2}), pointing directly over the TA Hotspot (Figure~\ref{fig:grid}). M82 is the closest starburst galaxy to our galaxy. 

If the source is assumed to be at the same distance to M82 (3.7~Mpc) with a pure proton emission, the average coherent magnetic field perpendicular to the FOV required to cause this deflection is $B$$=$$11$~nG. The large deviations from the linear fit of Figure~\ref{fig:wedgescat2} imply that, in this region, the random field deflections have a large correlation length scale ($L_c$), and the random field ($B_{rms}$) of Equation \ref{eq:randdef} is on the same order of magnitude as the coherent field deflection.

Random variations of the data are created to estimate the uncertainty on the location of the source of the maximum significance energy-angle correlation. The energies of events outside the wedge are scrambled with other events outside the wedge. Inside, the energies of wedge events with an energy less than the wedge threshold, E$<$35 EeV, are randomized within the wedge. The locations of the 75 data events in the wedge with E$\geq$35~EeV are not changed. This ensures that the spectrum is not changed, inside or outside the wedge, and that the number of events E$\geq$35~EeV does not dramatically increase due to the Coldspot (\cite{Abbasi2018E}). The analysis, including scanning for maximum significance wedge correlations at all grid points, was repeated for 5000 of these random variations on the data.

The estimated location of each randomized data sets source is the most significant negative correlation near the known source grid point. The maximum distance searched, within a spherical cap centered on the known grid point, is the distance that minimizes the average $\tau$ inside the cap (correlations are more positive outside). A spherical cap limiter is necessary due to the fact that an entirely different set of events, than the wedge of interest, say on the other side of the FOV, can easily have a more significant correlation due to the number of scans done at each grid point. 

\begin{figure}[h]
\centering
    \includegraphics[width=1\linewidth]{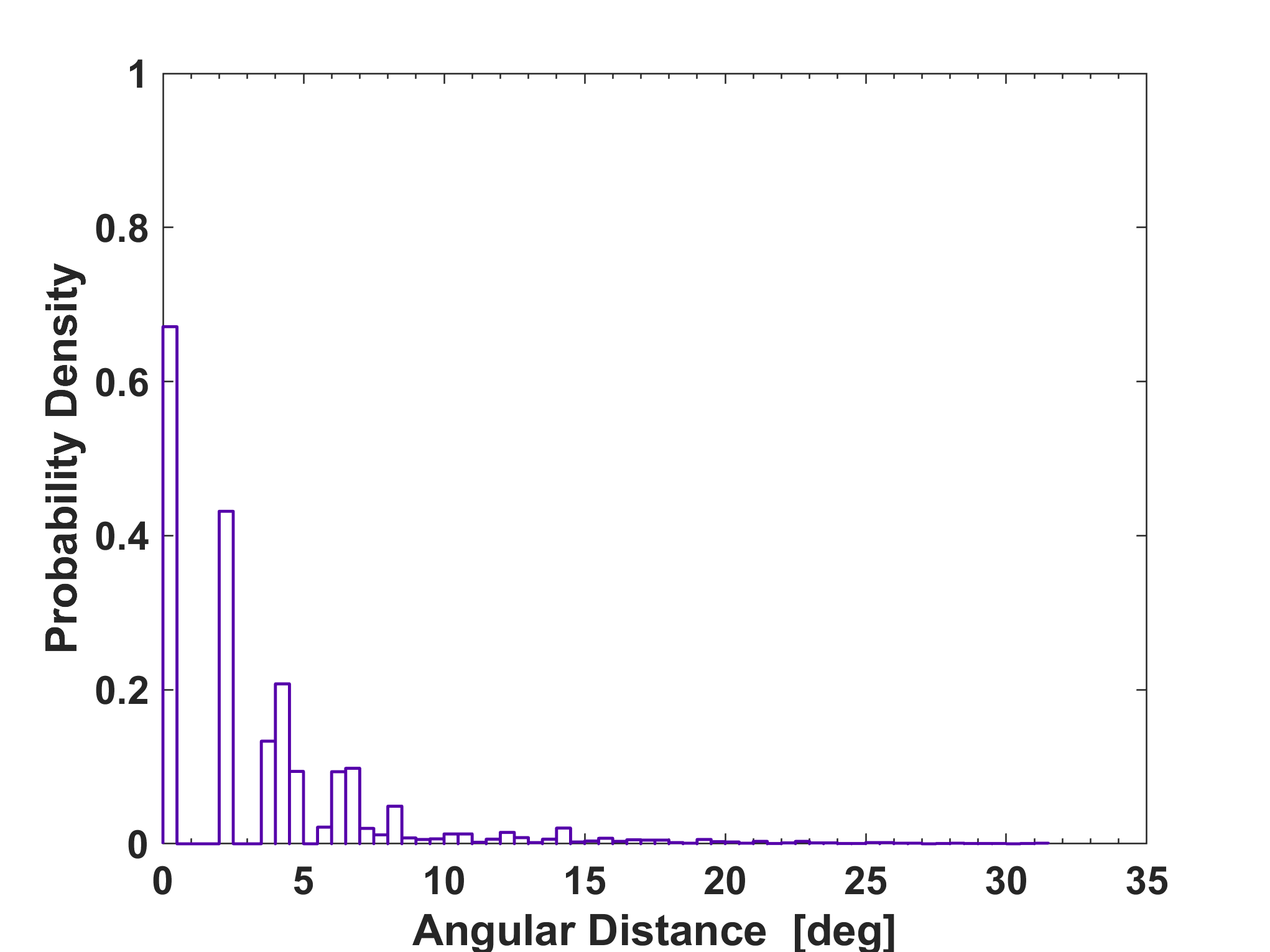}
  \caption{Distribution of distances from the actual most significant correlation grid point to those found in randomized data with the wedge embedded.}\label{fig:wedgedist}
\end{figure}

The result is that the apparent sources have a median distance of 2.4$\Deg$ from the original source with a $+1\sigma$ quantile of 6.8$\Deg$ and 6.2$\%$ are greater than or equal to 11.3$\Deg$ away (the angular distance from M82 to the maximum significance grid point). The distribution of distances is shown in Figure~\ref{fig:wedgedist}. This distribution means that M82 is not excluded as a possible source of the events in this energy-angle correlation and the TA UHECR Hotspot/Coldspot \cite{Abbasi2018E}.

The result presented here appears to be consistent with the results of \cite{He:2014mqa} that used a Bayesian analysis of the relative deflection of TA Hotspot events in two energy bins (E$<$75~EeV and E$>$75~EeV). Their result was a 99.8$\%$ probability that M82 is the hotspot source.

\begin{figure}[h]
\centering
    \includegraphics[width=1\linewidth]{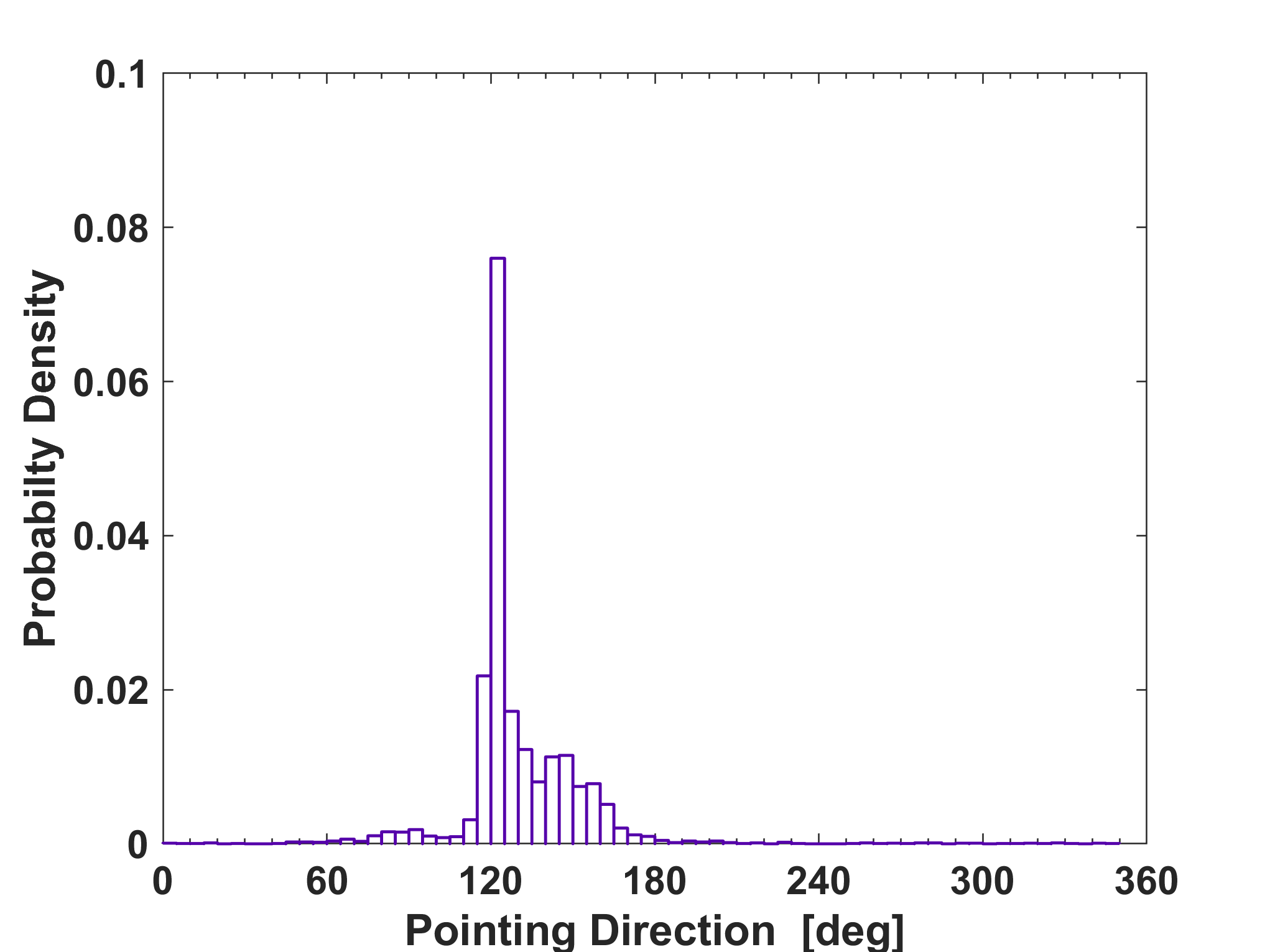}
  \caption{Distribution of pointing direction of wedges found in randomized data with the most significant wedge embedded.}\label{fig:wedgedir}
\end{figure}

According to the recent light polarization measurement of M82's magnetic field in \cite{Jones_2019}, the integrated magnetic field angle is 351$\Deg$ in equatorial coordinates using the same definition as Section~\ref{ssec:wedge}. Rotating into supergalactic coordinates results in an angle of 308$\Deg$. The coherent magnetic field direction necessary to create the most significant multiplet is 120$\pm$90$\Deg$, so it is either 82$\Deg$ or 98$\Deg$ from M82's magnetic field direction. The circular standard deviation of the pointing direction of the wedge simulations shown in Figure \ref{fig:wedgedir} is 21$\Deg$, which means the wedge magnetic field direction is $\geq$4$\sigma$ different from M82's magnetic field direction. This direction discrepancy implies that if M82 is the source then magnetic fields outside M82 were the primary source of multiplet pattern deflections.

\subsection{Supergalactic Field Estimate}
The average linear fit to $1/E$ versus angular distance from the grid point, inside wedges with a negative correlation, can give an estimate of coherent magnetic field strength times distance traveled through the field (see Equation~\ref{eq:deflect} as shown in Figure~\ref{fig:wedgescat1}). These $B$$\times$$S$ values are independent of the ranked correlation pre-trial significances, which were maximized to choose the wedge parameters. 

\begin{figure}[h]
\centering
    \includegraphics[width=1\linewidth]{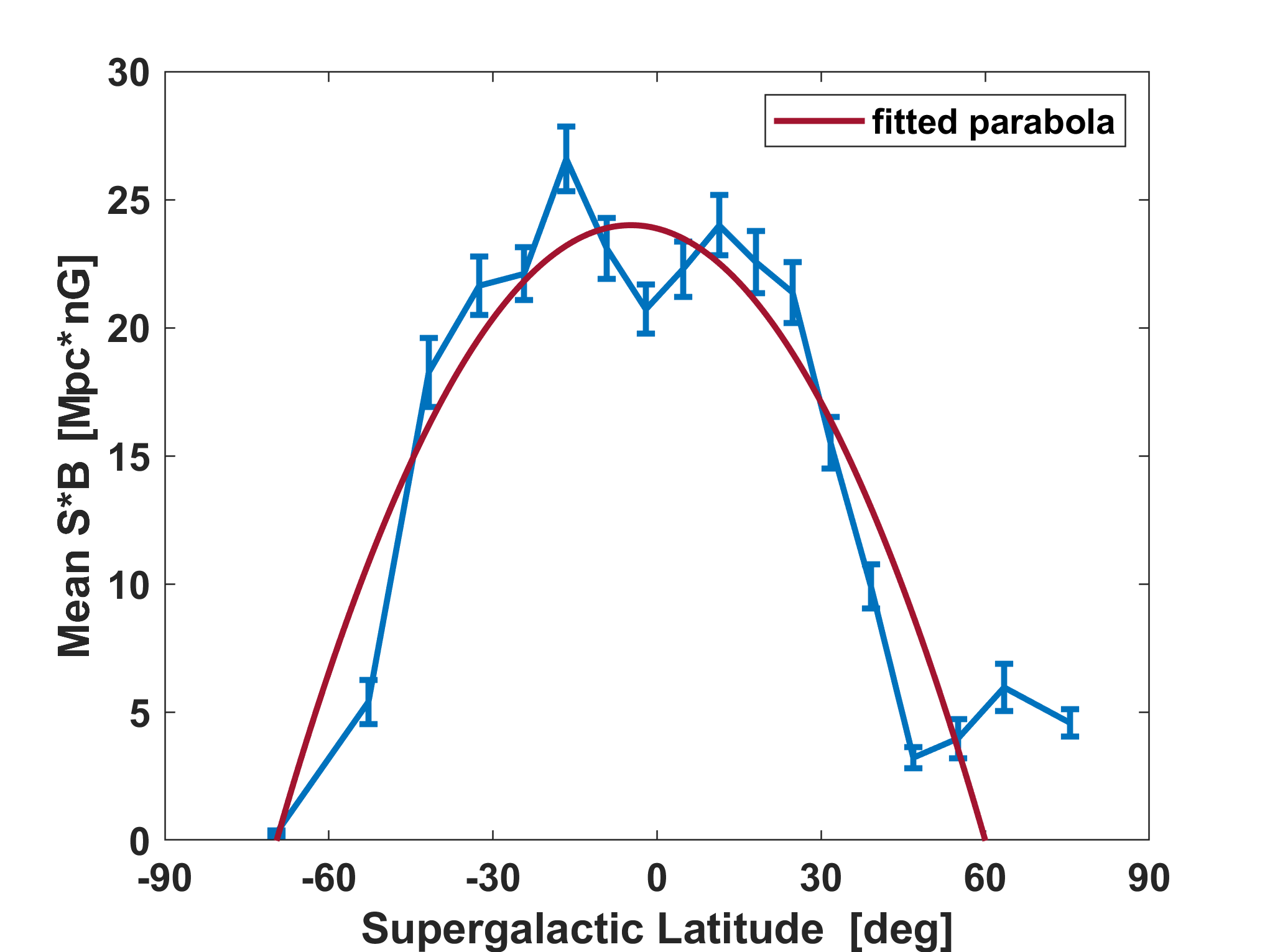}
  \caption{Mean $B$$\times$$S$ inside equal solid angle SGB bins setting $B$$\times$$S$$=0$ for wedges with positive correlations. The fitted parabola also demonstrates the correlation between apparent magnetic deflection multiplets with the supergalactic plane. These values are independent of the ranked correlation pre-trial significances, which were maximized to choose the wedge parameters. The mean within $\lvert SGB\rvert\leq40\Deg$ is $<$$B$$\times$$S$$>$ = 21~nG*Mpc. If proton is the assumed composition, then the average coherent field component, perpendicular to the FOV, in the vicinity of the supergalactic plane assuming a distance of 3.7~Mpc is 5.6~nG.}\label{fig:meanSB_zero}
\end{figure}

If the coherent magnetic field in the vicinity of positive correlations is considered negligible, and those correlations are set to $B$$\times$$S$ $=0$, then the mean $B$$\times$$S$ in supergalactic latitude (SGB) bins appears as Figure~\ref{fig:meanSB_zero}. Given $<$$B$$\times$$S$$>$~=~21~nG*Mpc and if the composition is protonic, then the average coherent field component, perpendicular to the FOV, in the vicinity of the supergalactic plane ($\lvert SGB\rvert\leq40\Deg$) is 5.6~nG (assuming a source distance of 3.7~Mpc).

\begin{figure}[h]
\centering
    \includegraphics[width=1\linewidth]{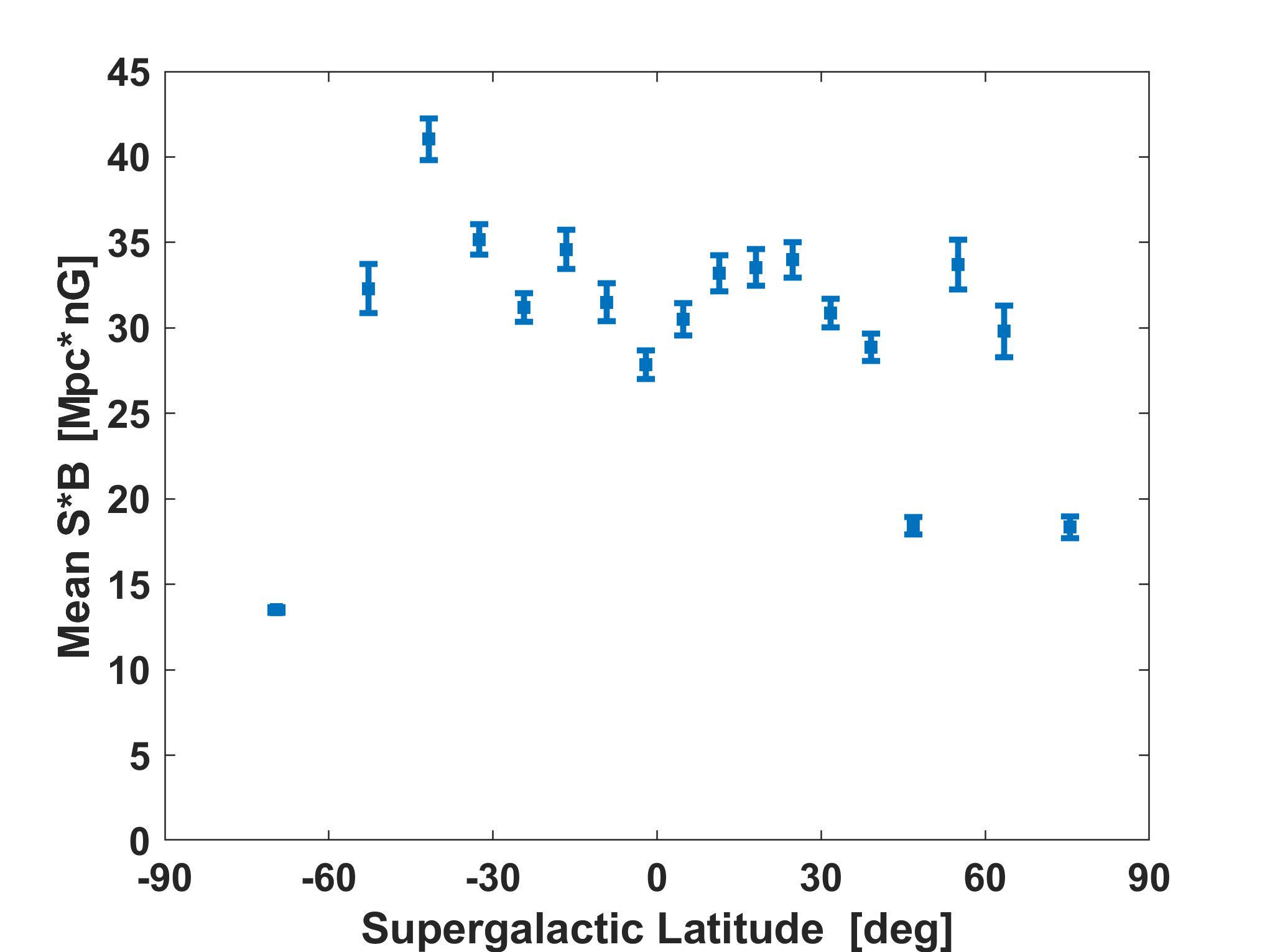}
  \caption{Mean $B$$\times$$S$ inside equal solid angle SGB bins not counting wedges with positive correlations. These values are independent of the ranked correlation pre-trial significances, which were maximized to choose the wedge parameters. The mean within $\lvert SGB\rvert\leq40\Deg$ is $<$$B$$\times$$S$$>$ = 32~nG*Mpc. If proton is the assumed composition, then the average coherent field component, perpendicular to the FOV, in the vicinity of the supergalactic plane assuming a distance of 3.7~Mpc is 8.6~nG.}\label{fig:meanSB_nan}
\end{figure}

If the coherent magnetic field in the vicinity of positive correlations is considered to be unknown, and those correlations are ignored, then the mean $B$$\times$$S$ in supergalactic latitude (SGB) bins appears as Figure~\ref{fig:meanSB_nan}. If proton is the assumed composition, then the average coherent field component, perpendicular to the FOV, in the vicinity of the supergalactic plane assuming a distance of 3.7~Mpc is $\sim$8.6~nG.

Recently in \cite{Globus2019OnTO}, the best fit average extragalactic field to the PAO dipole, assuming a local large scale structure (LSS) distribution of sources according to CosmicFlows-2 catalog, was estimated to be 0.6 nG using the PAO mixed composition E$\geq$8 EeV in \cite{Aab:2017njo}. If the TA composition is largely protonic, as in \cite{Abbasi2018Comp}, then the average distance traveled, $S$, necessary for agreement with PAO on the extragalactic field strength is $\sim$50 Mpc. The mean distance of galaxies in the CosmicFlows-2 catalog, within the GZK horizon of $\sim$100 Mpc, is 51 Mpc \cite{2013AJ....146...86T}. Given the model and experimental uncertainties, TA and PAO seem to have a good order of magnitude agreement on the extragalactic field strength.

\section{Systematic Checks}\label{sec:systematics}
A test of variation of isotropic MC parameters was done by calculating the significance of the supergalactic structure for seven years of data using two different MC. The first MC used the actual positions of the data and randomized energies according to the energy spectrum. This result had a 4.3$\pm 0.2 \sigma$ significance (two out of 200,000 trials with an $a$ parameter greater than data) and was reported in \cite{LundquistUHECR2019}. That significance is consistent with the current result, using completely isotropic position MC, of 4.2$\sigma$ with over five times more MC sets used in the calculation.

\subsection{Energy/Temperature Systematic}
Though ranked correlation is likely to decrease the effect of systematics, and temperature is taken into account for energy reconstruction, there is a possibility of a residual amount of correlation between the two. To test for this, each event trigger time was assigned the closest in time temperature measurement from three Delta, Utah stations taken from the NOAA databases \cite{tempdata}. 

Using the ten years of data set, the correlation between energy and temperature is $\tau=0.027$ (a small tendency for energy to increase with increasing temperature) with a 0.9$\%$ probability it is actually zero given enough samples. Additionally, there may be a correlation between angular distance from a grid point, and temperature as the average temperature in equatorial Right Ascension (R.A.) varies about 5$\Deg$. 

To check the possibility that the supergalactic structure found could be an artifact of temperature variations the partial Kendall correlation, $\tau_{xy.z}$, between energy and angular distance is done, removing temperature as a possible confounding variable. This is shown in Equation~\ref{eq:partcorr} ($x$ stands for energy, $y$ for angular distance, and $z$ for temperature).

\begin{equation}
\tau_{xy.z} = \frac{\tau_{xy}-\tau_{xz}\tau_{yz}}{(1-\tau_{xz}^2)(1-\tau_{yz}^2)}
\label{eq:partcorr}
\end{equation}

The average $\tau_{xy.z}$ in equal solid angle bins of supergalactic latitude (SGB) results in a parabolic fit curvature decrease of 0.8$\%$. This decrease is a very small difference and likely an effect of noise in the temperature measurements used. Therefore, no evidence for a temperature anisotropy producing the results is found.

\subsection{Galactic Field Influence}
The energy-angle correlation wedge parameter space should minimize the number of exclusively galactic field created correlations that result from the correlation significance scan. The minimum wedge distance is 15$\Deg$ (with a resulting mean of 63$\Deg$ for 10 years of data), and the minimum wedge width is 10$\Deg$ (with a mean of 26$\Deg$). The mean galactic magnetic field deflection expectation for UHECR protons with energies E$\geq$26~EeV (the average data wedge energy threshold) is $\leq\sim$15$\Deg$ for the various models in \cite{Farrar_2019}, and the expected dispersion around the mean is $<$10$\Deg$ for E$\geq$10~EeV.
 
\begin{figure}[h]
\centering
  \subfigure[]{%
   \includegraphics[width=.5\textwidth]{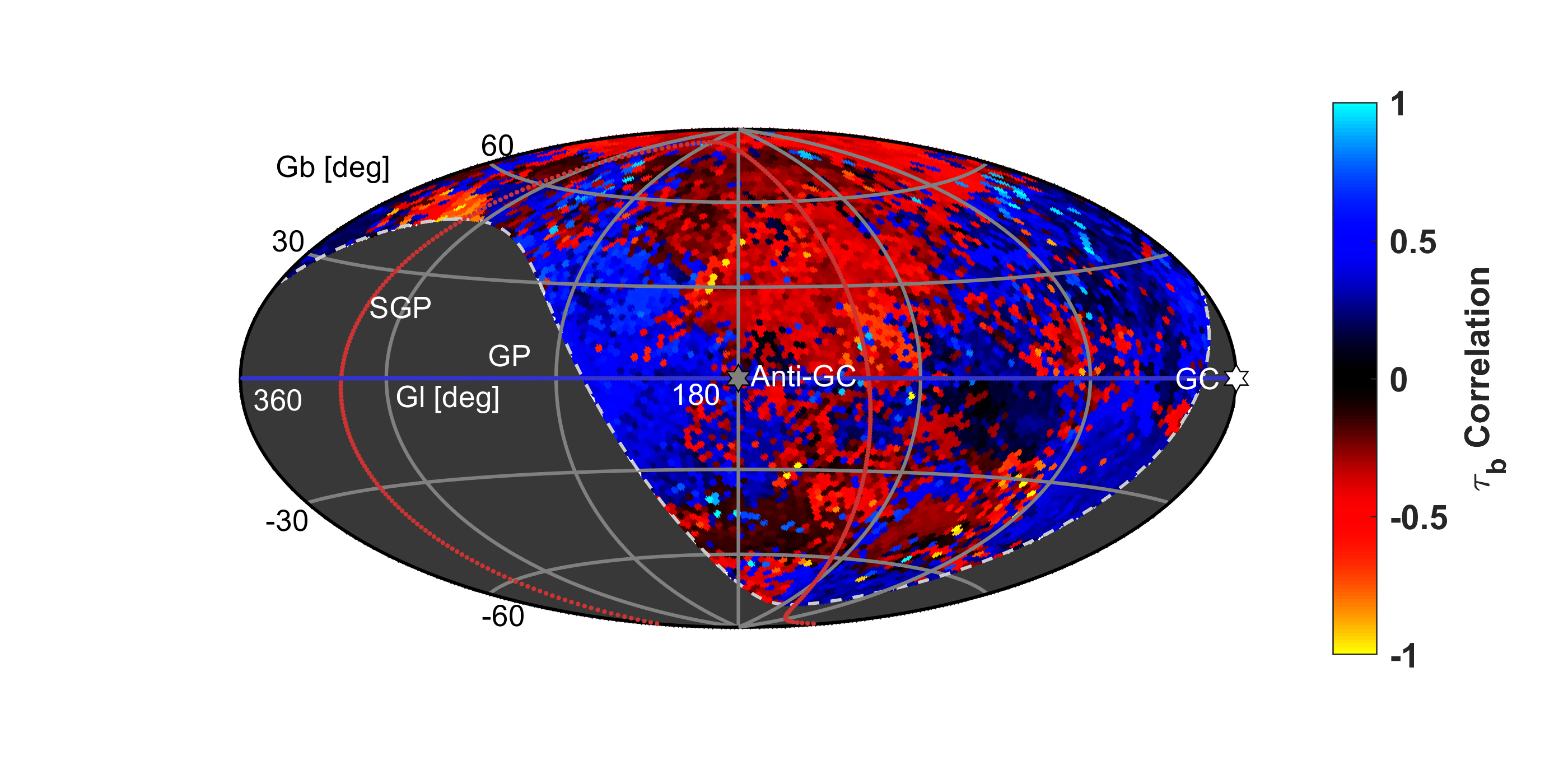}\label{fig:corrtau10gal}}
   \subfigure[]{%
    \includegraphics[width=.5\textwidth]{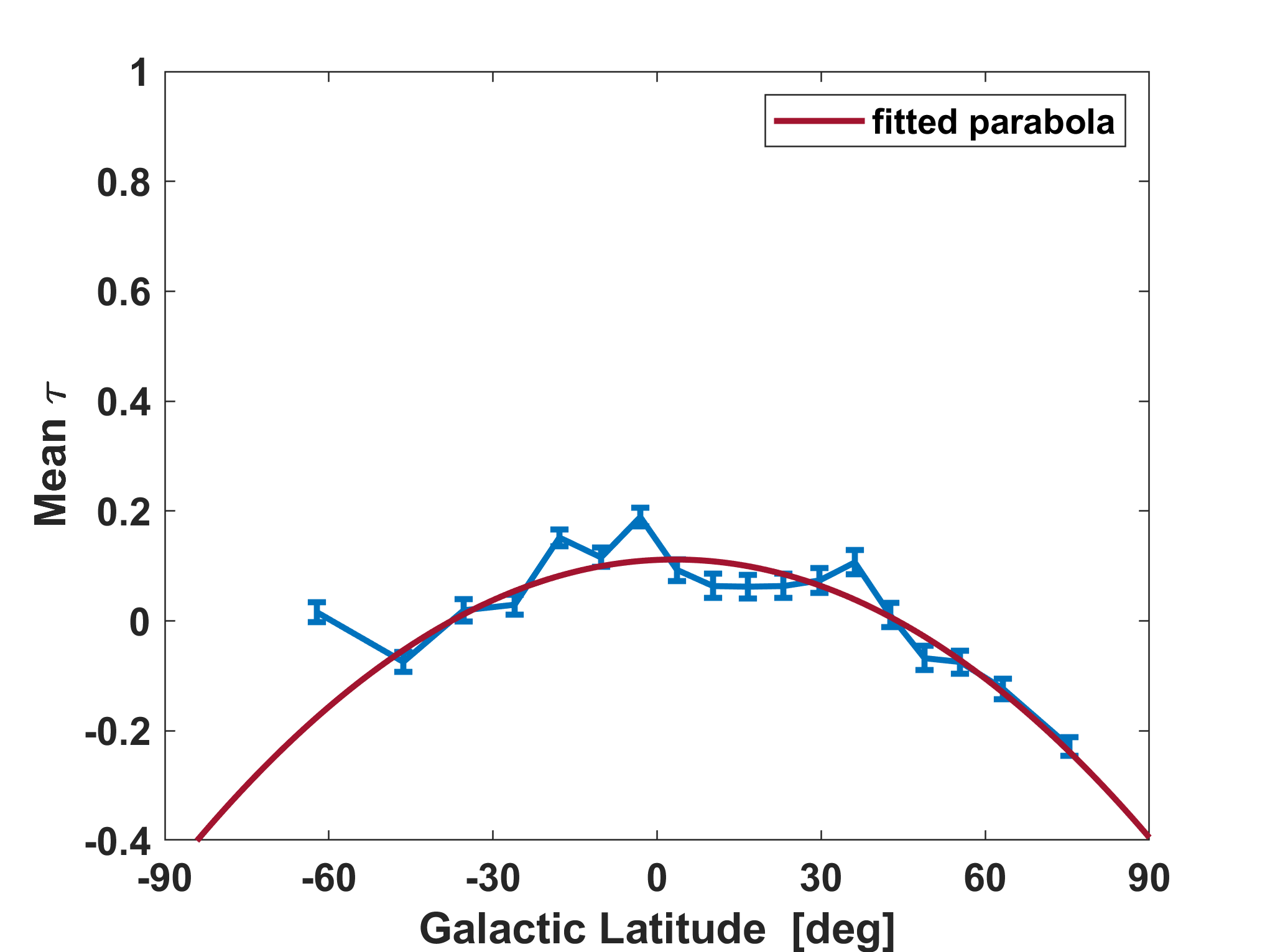}\label{fig:Galave10}}
  \caption{Ten years of data result shown in galactic coordinates. (a) Hammer-Aitoff galactic projection of the correlation strength $\tau$ for all grid points. Negative correlations expected for magnetic deflections are not apparent around the galactic plane. (b) Mean $\tau$ inside equal solid angle bins of galactic latitude (Gb). The resulting correlation structure curvature is $a$$=$$-6.7$$\times$$10^{-5}$.}
\end{figure}

\begin{figure}[h]
\centering
   \includegraphics[width=1\linewidth]{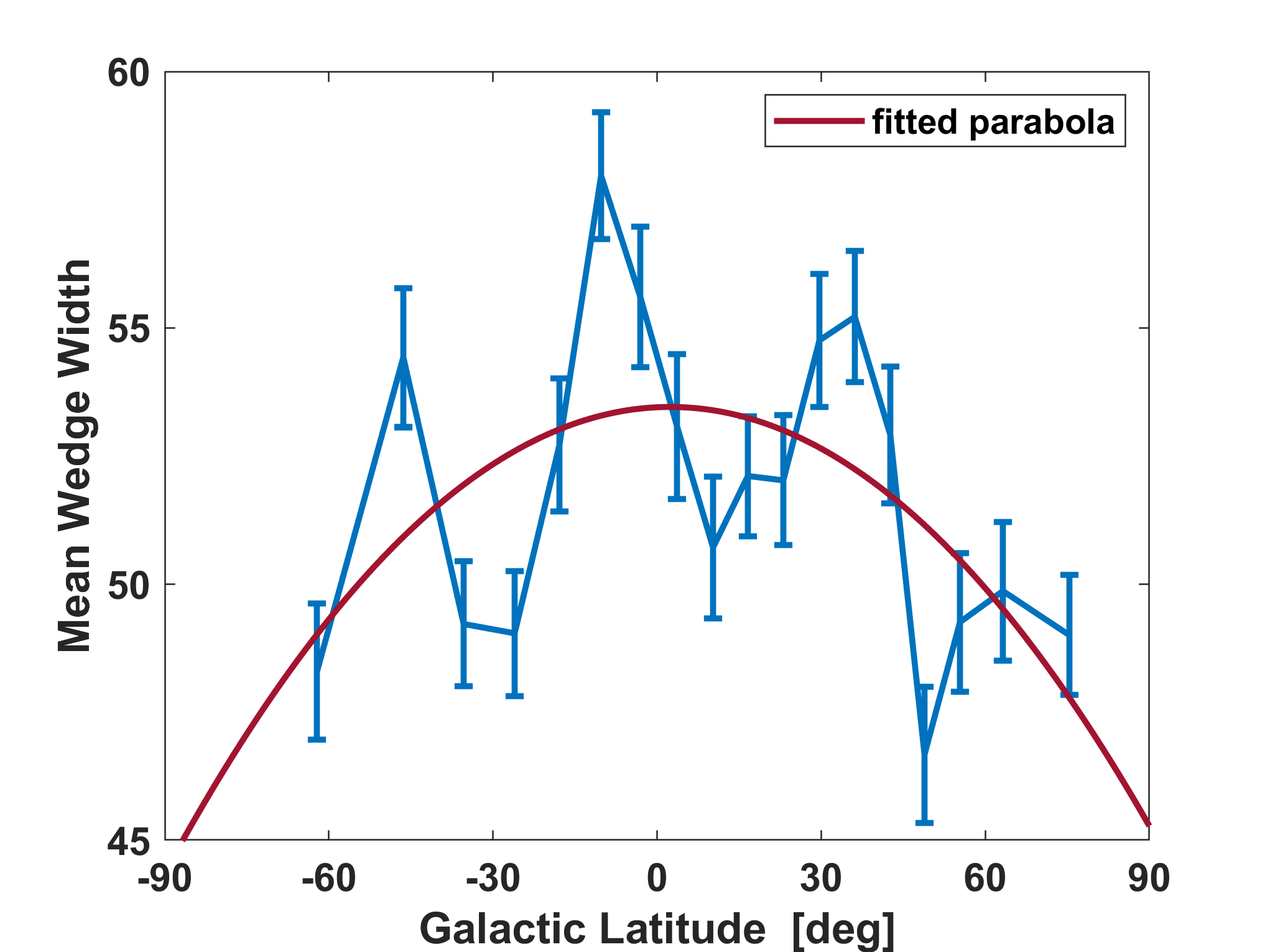}
  \caption{The mean wedge width inside equal solid angle bins of galactic latitude (Gb) for the ten years of data result. Wider bins are consistent with larger random field deflections near the galactic plane.}\label{fig:galwidths}
\end{figure}

The result of rotating into galactic coordinates and plotting the $\tau$ at each grid point for the ten years of data is shown in Figure~\ref{fig:corrtau10gal}. The negative curvature of the average $\tau$ with respect to galactic latitude (Gb), shown in Figure~\ref{fig:Galave10}, could suggest that possible magnetic deflections from apparent sources closer to the galactic plane are influenced by galactic magnetic fields with different directions than the average extragalactic fields. This behavior is consistent with the average widening of the wedge bins near the galactic plane shown in Figure~\ref{fig:galwidths}.

\begin{figure}[h]
\centering
   \includegraphics[width=1\linewidth]{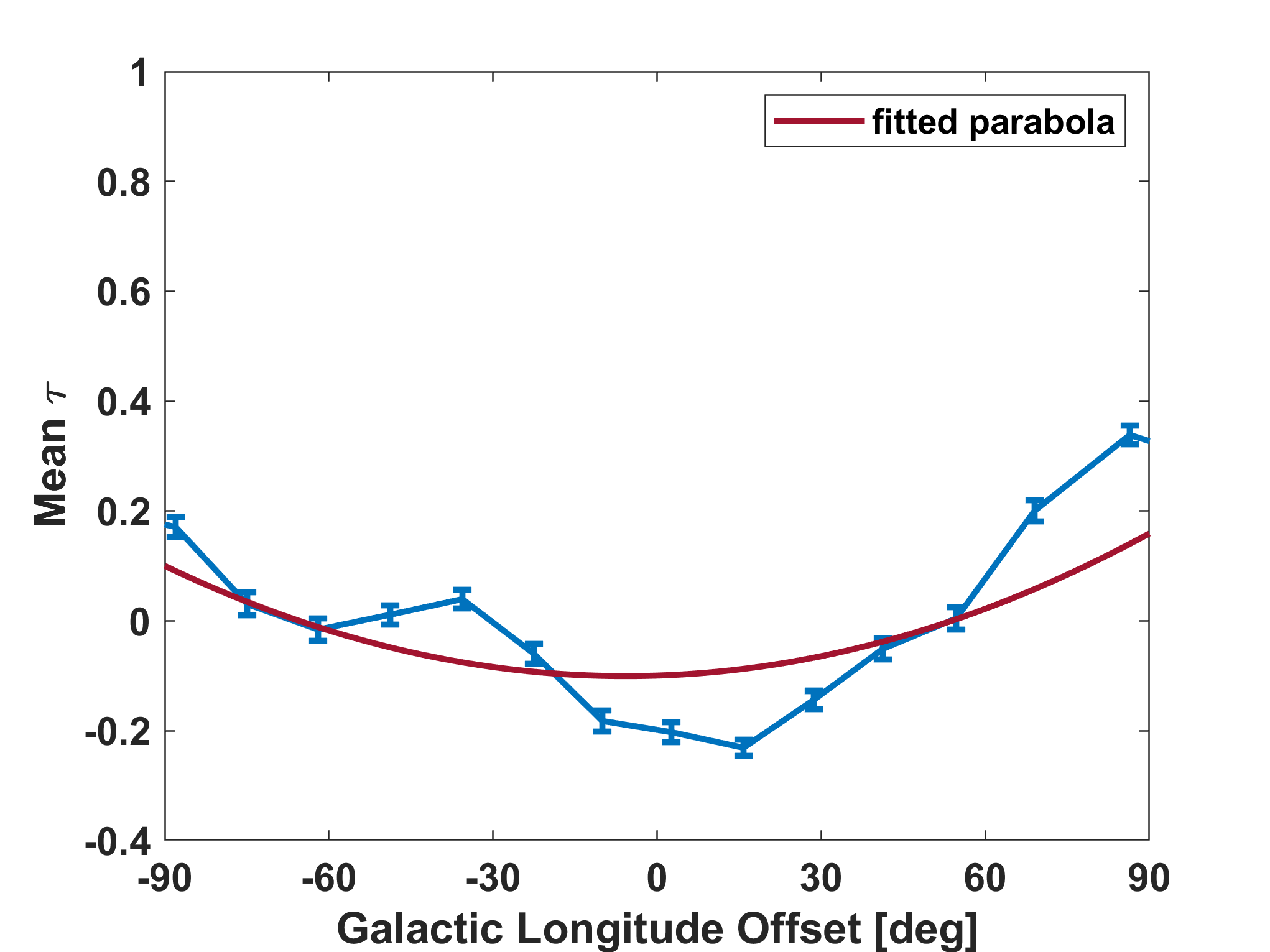}
  \caption{Ten years of data result shown in galactic coordinates for the mean $\tau$ inside equal solid angle bins of galactic longitude (Gl) centered on the intersection between the galactic plane (GP) and the supergalactic plane (SGP).}\label{fig:GalRotAve}
\end{figure}

Additionally, no apparent galactic structure of multiplets is found by the method of Section \ref{sec:sigmethod} when rotating the galactic coordinates by 90$\Deg$. This is shown in Figure~\ref{fig:GalRotAve} by the average $\tau$ in equal solid angle bins of galactic longitude (Gl) centered on the intersection between the galactic plane (GP) and the supergalactic plane (SGP). This rotation is where the correlations appear to have the most galactic symmetry according to Figure~\ref{fig:corrtau10gal} though the resulting correlation curvature $a$ from the fit is 18$\%$ of the supergalactic curvature result.

\section{Summary}\label{sec:summary}

Intermediate-scale energy-angle correlations inside spherical cap sections, or ``wedges,'' have been shown to have a $\sim$4$\sigma$ correlation with the supergalactic plane. Seven years of Telescope Array (TA) data has a 4.2$\sigma$ post-trial significance, and the ten years of data significance is 4.1$\sigma$ post-trial. These results may be evidence of large scale extragalactic magnetic diffusion of UHECR from sources within the local large scale structure as there does not appear to be a galactic correlation structure. 

Additionally, the highest significance single energy-angle correlation has increased from a pre-trial 4.6$\sigma$ significance (in the seven years of data) to 5.1$\sigma$ (in the ten years of data) with no new scan of wedge parameters. This correlation lies directly over the TA Hotspot, and its origin point is consistent with the starburst galaxy M82 being a source of these events. This result is consistent with other results assuming magnetic deflection such as \cite{He:2014mqa} and with the starburst galaxy overdensity anisotropy study of \cite{Aab:2018chp}.

If M82 is the source of the most significant correlation, then the average coherent magnetic field component perpendicular to the FOV, within this section of the sky, is estimated to be 11 nG assuming a purely proton composition. 

The average perpendicular magnetic field correlated with the supergalactic plane is also estimated to be on the order of ~10 nG assuming a cosmic ray travel length of 3.7 Mpc and a proton composition. A mixed composition and/or a longer travel length results in a smaller magnetic field estimate. This result is consistent with other estimates of extragalactic magnetic fields via theory, simulation and astrophysical measurements (\cite{Ryu:1998up}, \cite{Globus2019OnTO}, \cite{1994RPPh...57..325K} for example).

Confirmation of these results awaits sufficient data to be collected by the TA expansion to TAx4 \cite{Kido:2019enj}.

\acknowledgments

The Telescope Array experiment is supported by the Japan Society for the Promotion of Science(JSPS) through Grants-in-Aid for Priority Area 431, Specially Promoted Research JP21000002, Scientific Research (S) JP19104006, Specially Promoted Research JP15H05693, Scientific Research (S) JP15H05741, Science Research (A) JP18H03705, and for Young Scientists (A) JPH26707011; by the joint research program of the Institute for Cosmic Ray Research (ICRR), The University of Tokyo; by the U.S. National Science Foundation awards PHY-0601915, PHY-1404495, PHY-1404502, and PHY-1607727; by the National Research Foundation of Korea
(2016R1A2B4014967, 2016R1A5A1013277, \\
2017K1A4A3015188, 2017R1A2A1A05071429);
by the Russian Academy of Sciences, RFBR grant 20-02-00625a (INR), IISN project No. 4.4502.13, and Belgian Science Policy under IUAP VII/37 (ULB). The foundations of Dr. Ezekiel R. and Edna Wattis Dumke, Willard L. Eccles, and George S. and Dolores Dor\'e Eccles all helped with generous donations. The State of Utah supported the project through its Economic Development Board, and the University of Utah through the Office of the Vice President for Research. The experimental site became available through the cooperation of the Utah School and Institutional Trust Lands Administration (SITLA), U.S. Bureau of Land Management (BLM), and the U.S. Air Force. We appreciate the assistance of the State of Utah and Fillmore offices of the BLM in crafting the Plan of Development for the site. Patrick Shea assisted the collaboration with valuable advice on a variety of topics. The people and the officials of Millard County, Utah have been a source of steadfast and warm support for our work which we greatly appreciate. We are indebted to the Millard County Road Department for their efforts to maintain and clear the roads which get us to our sites. We gratefully acknowledge the contribution from the technical staffs of our home institutions. An allocation of computer time from the Center for High Performance Computing at the University of Utah is gratefully acknowledged.

\bibliographystyle{aasjournal}
\bibliography{MyThesis3}

\end{document}